\documentclass[journal,final,letterpaper]{IEEEtran}
\usepackage{cite}
\usepackage[dvipdf]{graphicx}

\usepackage[cmex10]{amsmath}
\usepackage{color}

\newtheorem{defn}{Definition}

\newtheorem{thm}[defn]{Theorem}
\newtheorem{algo}{Algorithm}

\begin{document}
\title{Path Gain Algebraic Formulation for the Scalar Linear Network Coding Problem}
\author{Abhay~T.~Subramanian and
        Andrew~Thangaraj,~\IEEEmembership{Member, IEEE}
\thanks{A. Thangaraj is with the Department of Electrical Engineering, Indian Institute of Technology (IIT) Madras, Chennai, India. E-mail: {andrew@iitm.ac.in}}
\thanks{A. T. Subramanian was a master's student at IIT Madras at the time of this work. He is currently a doctoral student at the Department of Management Science and Engineering in Stanford University. Email: {abhayts@stanford.edu}}}


\maketitle

\begin{abstract}
In the algebraic view, the solution to a network coding problem is seen as a variety specified by a system of polynomial equations typically derived by using edge-to-edge gains as variables. The output from each sink is equated to its demand to obtain polynomial equations. In this work, we propose a method to derive the polynomial equations using source-to-sink path gains as the variables. In the path gain formulation, we show that linear and quadratic equations suffice; therefore, network coding becomes equivalent to a system of polynomial equations of maximum degree 2. We present algorithms for generating the equations in the path gains and for converting path gain solutions to edge-to-edge gain solutions. Because of the low degree, simplification is readily possible for the system of equations obtained using path gains. Using small-sized network coding problems, we show that the path gain approach results in simpler equations and determines solvability of the problem in certain cases. On a larger network (with 87 nodes and 161 edges), we show how the path gain approach continues to provide deterministic solutions to some network coding problems.
\end{abstract}
\begin{IEEEkeywords}
Algebraic network coding, Network coding, Scalar linear network coding.
\end{IEEEkeywords}
\section{Introduction}
\IEEEPARstart{T}{he} idea of network coding over error-free networks, pioneered in \cite{ahlswede}, has been a subject of active current research. The general idea of linear network coding, where intermediate nodes linearly combine incoming packets, was explored in \cite{linnwc}. A simple and effective algebraic formulation of the general network coding problem was introduced in \cite{algebraic}. This established a direct connection between a network information flow problem and an algebraic variety over the closure of a finite field. 

Using the formulations of \cite{linnwc, algebraic}, the multicast network coding problem, where one source transmits at the same rate to a set of sinks, has been characterized almost completely. A linear network code exists for the multicast case in a large enough finite field and can be found in polynomial time \cite{jaggi}. The insufficiency of linear coding in the non-multicast case has been demonstrated in \cite{doughinsuff}. Recent work in \cite{doughpoly} and \cite{iisc} has shown the restrictions imposed on the field characteristic for the scalar linear solvability of a general network coding problem. See \cite{doughpoly} for more non-multicast examples.

In the algebraic view, the network code is seen as a variety specified by a system of polynomial equations in multiple variables taking values from a finite field \cite{algebraic}\cite{doughpoly}. To derive the equations corresponding to a given network coding problem, edge-to-edge gains are assigned as variables. For every node, the flow on outgoing edges is written down in terms of the flows on the incoming edges using the edge-to-edge gains. The flow propagates in this manner from the sources to the sinks. The output from the sink is equated to its demand, and polynomial equations in the edge-to-edge gains are obtained. 

In this work, we propose a method to derive the equations using path gains as the variables. The gain on every source to sink path becomes a variable in the proposed formulation. In the method of \cite{algebraic}, the path gain would be a product of several edge-to-edge gain variables. The advantage of the path gain formulation is that the final equations are only linear and quadratic, as shown in the remainder of this article. Because of the low degree and the inherent nature of the scalar linear network coding problem, simplification is readily possible for the system of equations. We provide an algorithm to compute the equations in the path gain formulation, and demonstrate the efficacy of the path gain approach by illustrative examples. Starting with the butterfly network and other interesting small-sized network coding problems, we show that the path gain approach provides results on solvability of the problem. On a larger network (with 87 nodes and 161 edges), we show how the path gain approach continues to provide solutions to some network coding problems.

The path gain formulation is equivalent to the edge-to-edge gain formulation and can be derived from it. Therefore, the work presented in this article is a method to simplify the equations generated by the edge-to-edge gain variable assignment. While the number of variables in the edge-to-edge formulation is of the order of the number of edges, the number of monomial terms in these variables is exponential in the number of edges. Hence, the polynomial system is of size that can be exponential in the size of the network. Assigning variable names to the paths (which can be exponential in the size of the network in number) does not necessarily make the path gain formulation more complex than the edge-to-edge gain formulation as far as solving the equations is concerned. However, in an actual implementation, the edge-to-edge gains are to be used. To complete the path gain formulation, we provide an algorithm to compute the edge-to-edge gains from the path gains. 

Though there are several other standard methods to simplify systems of polynomial equations (such as Gr\"{o}bner basis methods), many problems in the area of solving systems of polynomial equations (and in network coding with multiple sources and sinks) are either NP-hard or undecidable. In this light, the path gain formulation appears to be simpler than the edge-to-edge formulation in the sense that simplifications and solutions are easier in several examples (both small and large). 

Several methods and techniques to study the network coding problem have been introduced by many researchers in this area that has seen intense recent research activity. Following the information-theoretic methods in \cite{ahlswede}, more information-theoretic methods were used for characterizing network coding for multimessage unicasts in \cite{4036066}. The algebraic formulation in \cite{algebraic} provided an elegant and powerful method to study network coding. Random network coding \cite{1705002}, which is a popular choice in practical implementations, was introduced and studied using algebraic tools. The linear programming formulation has seen applications in wireless network coding \cite{1532487} and optimizing network coding with a cost criterion \cite{1638546}. The combinatorial approach, proposed and developed in \cite{1638536} and \cite{1603756}, has provided methods for studying the field sizes in network coding problems.

In the context of the prior work cited above, the path gain formulation for algebraic network coding presents the equivalence between network coding and a maximum-degree-2 system of polynomial equations for the first time. The equivalence is achieved without introducing any new monomial terms that are not present in the original system. The equations obtained from the path gain formulation are amenable to considerable simplification in several cases of interest. Hence, the path gain method can provide deterministic solutions to several linear network coding problems. The method can, in some cases, provide results on solvability. The primary utility of the method is likely to be in larger examples. As an illustration, for the network (in Fig. \ref{isp}) with 87 nodes and 161 edges, we present results of solutions to certain network coding problems with multiple sources and sinks in Section \ref{sec:bigger-example}.

The rest of this article is organized as follows. We will start with a notational description of the network coding problem in Section~\ref{sec:nwc}, which also introduces the edge-to-edge gain algebraic formulation of \cite{algebraic}. The path gain formulation is presented in Section~\ref{sec:transform}, where we provide a graph transformation algorithm that is used to represent and compute the equations in a transformed graph. At the end of the section, we show how the equations derived from path gain variables are amenable to easy simplifications. In Section~\ref{sec:example}, we illustrate the advantages of the path gain formulation using various example networks drawn from the literature. We also provide results for a large Internet Service Provider (ISP) network. In Section~\ref{sec:reverse}, we give an algorithm (that uses the transformed graph) to derive the edge-to-edge gains from the path gains. Finally, we provide concluding remarks in Section \ref{sec:conclusion}.

\section{The Network Coding Problem}
\label{sec:nwc}
The communication network is modeled as a directed, acyclic multigraph, $G=(V,E)$, where the node set $V$ represents the terminals and switches in the network and the edge set $E$ represents the communication links. It is assumed that all communication links are error-free and have unit capacity. 

\subsection{Notation}
For a given edge $e = (u,v)$, we denote:
\begin{align*}
&u = \text{tail}(e)\\
&v = \text{head}(e)
\end{align*}

For each node $v \in V$ , we define
\begin{align*}
&I(v) = \{e \in E : \text{head}(e) = v\},\\
&O(v) = \{e \in E : \text{tail}(e) = v\}.
\end{align*}

Let us further assume the following without loss of generality:
\begin{enumerate}
\item A node $v$ is a source node iff $|I(v)| = 0$ and all source nodes produce exactly one unit of data per unit time.
\item A node $v$ is a sink node iff $|O(v)|=0$ and all sink nodes demand exactly one unit of data per unit time.
\end{enumerate}

In cases where a node $v$ produces (demands) more than one data symbol, we can add virtual source (sink) nodes that produce (demand) exactly one data symbol, have exactly one output (input) link connecting them to $v$ and no input (output) links.

Then, the set of source and sink nodes is defined as follows:
\begin{align*}
&S = \{v \in V : |I(v)| = 0\} = \{s_1,s_2,\ldots,s_{|S|}\}.\\
&T = \{v \in V : |O(v)| = 0\} = \{t_1,t_2,\ldots,t_{|T|}\}.
\end{align*}
Let the sink $t_j$ demand the $s(j)$-th source. For every source $s$, a virtual incoming edge $e(s)$ is added for notational convenience (as in the edge-to-edge gain formulation \cite{algebraic}).

Let us now assume that we use a finite alphabet $H$. For each edge $e$, an edge function is then defined as a mapping $\mathbf{f}_e : H^i \rightarrow H$, where $i = 1$ if $\text{tail}(e) \in S$ and $i = |I(\text{tail}(e))|$ otherwise. For a sink $t$, a virtual outgoing edge $e(t)$ is added to denote the output. The edge function on this virtual edge, which is a mapping denoted $\mathbf{f}_t: H^{|I(t)|}\rightarrow H$, is called the output function of the sink.

\begin{defn}
The collection of all the edge functions in a given network is defined as a {\it network code}. If all the edge functions are linear maps with respect to a field alphabet $H$, then the code is a {\it scalar linear code}.
\end{defn}

Let the data symbol generated at the $i$-th source node, $s_i \in S$, be denoted by $X_i$. The data symbol demanded by the $j$-th sink node, $t_j \in T$, is $X_{s(j)}$. The sources and sinks implicitly define a set of connection requirements for the given network $G$. The connection requirement is met at a sink $t_j$ if the output of the function $\mathbf{f}_{t_j}$ equals $X_{s(j)}$ for all inputs.

Given a network $G$, the set of source nodes $S$ and the set of sink nodes $T$, the network coding problem is to determine all the edge functions such that all the connection requirements are satisfied. If such a set of edge functions exists, then the network coding problem is {\it solvable}. If a set of linear edge functions (with respect to a finite field $H$) exists that satisfies all the connection requirements, then the network coding problem is {\it scalar-linearly solvable}.

In a scalar linear network coded flow (over a field $H$), the edge function of an edge $e$ can be written as $\sum_{i=1}^{|S|}a_iX_i$, where $a_i\in H$. We refer to $\sum_{i=1}^{|S|}a_iX_i$ as either the edge function of $e$ or the symbol flowing through $e$ and denote it as a vector $\mathbf{f}_e=[a_1\;a_2\;\cdots\;a_{|S|}]$. Similarly, the output function at the sink has a vector notation.

\subsection{Koetter-M\'{e}dard formulation and edge-to-edge gains}
\label{sec:km}
Solving the scalar linear network coding problem was formulated as a problem of solving a system of polynomial equations by Koetter and M\'{e}dard in \cite{algebraic}. The idea is to construct the linear edge function $\mathbf{f}_e$ for an edge $e=(u,v)$ recursively as follows:
\begin{equation}
  \label{eq:1}
\mathbf{f}_e=\sum_{e'\in I(u)}\alpha_{e',e}\mathbf{f}_{e'},  
\end{equation}
where $\alpha_{e',e}$ is an edge-to-edge gain. To start the recursion, the edges out of a source node $s_i$ are assigned the unit coding vector with a $1$ in the $i$-th position. The edge function for the remaining edges, found using (\ref{eq:1}), become vectors of polynomials in the edge-to-edge gains $\alpha_{e',e}$. Finally, at a sink $t_j$, the output edge function is equated to the unit vector with a $1$ in the $s(j)$-th position. These equations form a polynomial system in the edge-to-edge gains $\alpha_{e',e}$ for $e'\in I(\text{tail}(e))$.  

\begin{figure}[htb]
  \centering
\begin{picture}(0,0)%
\includegraphics{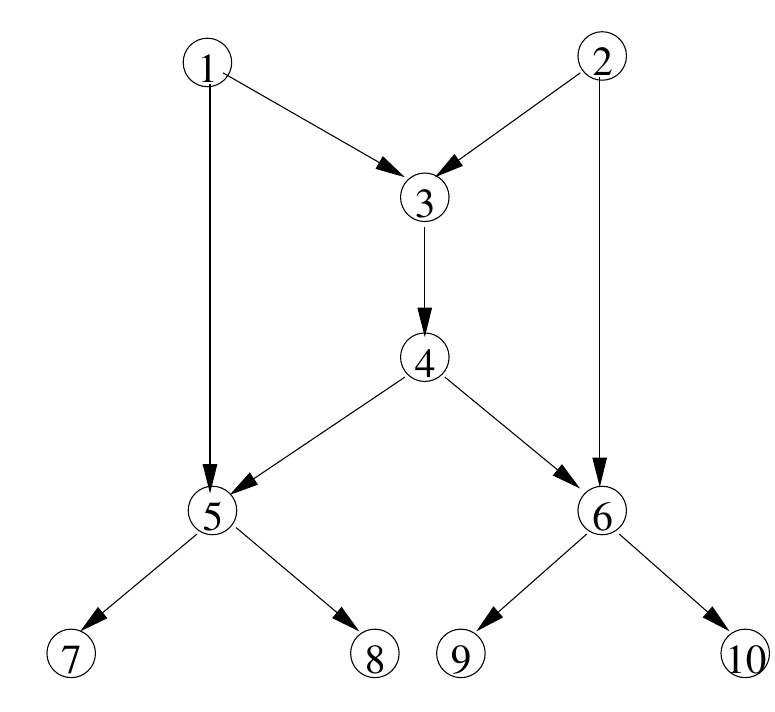}%
\end{picture}%
\setlength{\unitlength}{2735sp}%
\begingroup\makeatletter\ifx\SetFigFont\undefined%
\gdef\SetFigFont#1#2#3#4#5{%
  \reset@font\fontsize{#1}{#2pt}%
  \fontfamily{#3}\fontseries{#4}\fontshape{#5}%
  \selectfont}%
\fi\endgroup%
\begin{picture}(5346,4920)(706,-4531)
\put(2026,209){\makebox(0,0)[lb]{\smash{{\SetFigFont{7}{8.4}{\familydefault}{\mddefault}{\updefault}{\color[rgb]{0,0,0}$X_1$}%
}}}}
\put(4771,254){\makebox(0,0)[lb]{\smash{{\SetFigFont{7}{8.4}{\familydefault}{\mddefault}{\updefault}{\color[rgb]{0,0,0}$X_2$}%
}}}}
\put(2746,-376){\makebox(0,0)[lb]{\smash{{\SetFigFont{7}{8.4}{\familydefault}{\mddefault}{\updefault}{\color[rgb]{0,0,0}$X_1$}%
}}}}
\put(4051,-376){\makebox(0,0)[lb]{\smash{{\SetFigFont{7}{8.4}{\familydefault}{\mddefault}{\updefault}{\color[rgb]{0,0,0}$X_2$}%
}}}}
\put(2206,-1366){\makebox(0,0)[lb]{\smash{{\SetFigFont{7}{8.4}{\familydefault}{\mddefault}{\updefault}{\color[rgb]{0,0,0}$X_1$}%
}}}}
\put(4591,-1366){\makebox(0,0)[lb]{\smash{{\SetFigFont{7}{8.4}{\familydefault}{\mddefault}{\updefault}{\color[rgb]{0,0,0}$X_2$}%
}}}}
\put(3691,-2536){\makebox(0,0)[lb]{\smash{{\SetFigFont{7}{8.4}{\familydefault}{\mddefault}{\updefault}{\color[rgb]{0,0,0}$\alpha_1X_1+\alpha_2X_2$}%
}}}}
\put(5806,-4471){\makebox(0,0)[lb]{\smash{{\SetFigFont{7}{8.4}{\familydefault}{\mddefault}{\updefault}{\color[rgb]{0,0,0}$X_2$}%
}}}}
\put(3826,-4471){\makebox(0,0)[lb]{\smash{{\SetFigFont{7}{8.4}{\familydefault}{\mddefault}{\updefault}{\color[rgb]{0,0,0}$X_1$}%
}}}}
\put(3196,-4471){\makebox(0,0)[lb]{\smash{{\SetFigFont{7}{8.4}{\familydefault}{\mddefault}{\updefault}{\color[rgb]{0,0,0}$X_2$}%
}}}}
\put(1081,-4471){\makebox(0,0)[lb]{\smash{{\SetFigFont{7}{8.4}{\familydefault}{\mddefault}{\updefault}{\color[rgb]{0,0,0}$X_1$}%
}}}}
\put(3061,-1546){\makebox(0,0)[lb]{\smash{{\SetFigFont{7}{8.4}{\familydefault}{\mddefault}{\updefault}{\color[rgb]{0,0,0}$\alpha_1X_1+\alpha_2X_2$}%
}}}}
\put(2251,-2671){\makebox(0,0)[lb]{\smash{{\SetFigFont{7}{8.4}{\familydefault}{\mddefault}{\updefault}{\color[rgb]{0,0,0}$\alpha_1X_1+\alpha_2X_2$}%
}}}}
\put(721,-3706){\makebox(0,0)[lb]{\smash{{\SetFigFont{7}{8.4}{\familydefault}{\mddefault}{\updefault}{\color[rgb]{0,0,0}$+\alpha_4(\alpha_1X_1+\alpha_2X_2)$}%
}}}}
\put(946,-3526){\makebox(0,0)[lb]{\smash{{\SetFigFont{7}{8.4}{\familydefault}{\mddefault}{\updefault}{\color[rgb]{0,0,0}$\alpha_3X_1$}%
}}}}
\put(4096,-3616){\makebox(0,0)[lb]{\smash{{\SetFigFont{7}{8.4}{\familydefault}{\mddefault}{\updefault}{\color[rgb]{0,0,0}$\alpha_8X_2$}%
}}}}
\put(2431,-3301){\makebox(0,0)[lb]{\smash{{\SetFigFont{7}{8.4}{\familydefault}{\mddefault}{\updefault}{\color[rgb]{0,0,0}$\alpha_6(\alpha_1X_1+\alpha_2X_2)$}%
}}}}
\put(2611,-3481){\makebox(0,0)[lb]{\smash{{\SetFigFont{7}{8.4}{\familydefault}{\mddefault}{\updefault}{\color[rgb]{0,0,0}$+\alpha_5X_1$}%
}}}}
\put(3556,-3796){\makebox(0,0)[lb]{\smash{{\SetFigFont{7}{8.4}{\familydefault}{\mddefault}{\updefault}{\color[rgb]{0,0,0}$+\alpha_7(\alpha_1X_1+\alpha_2X_2)$}%
}}}}
\put(5356,-3616){\makebox(0,0)[lb]{\smash{{\SetFigFont{7}{8.4}{\familydefault}{\mddefault}{\updefault}{\color[rgb]{0,0,0}$+\alpha_{10}X_2$}%
}}}}
\put(4771,-3481){\makebox(0,0)[lb]{\smash{{\SetFigFont{7}{8.4}{\familydefault}{\mddefault}{\updefault}{\color[rgb]{0,0,0}$\alpha_9(\alpha_1X_1+\alpha_2X_2)$}%
}}}}
\end{picture}%
  \caption{Flow in the butterfly network.}
  \label{fig:bflow}
\end{figure}
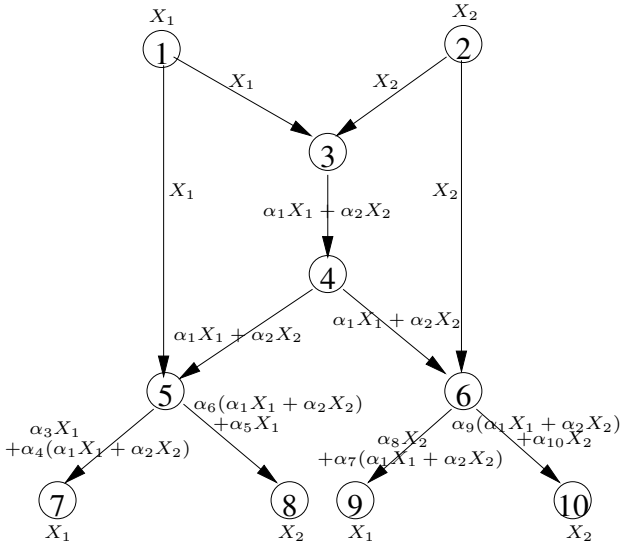

The Koetter-M\'{e}dard algebraic formulation is illustrated for the case of the modified butterfly network shown in Fig.~\ref{fig:bflow} with two sources and four sinks. Note that the network in Fig. \ref{fig:bflow} is identical to the classic butterfly network under our definition of sources and sinks. The edge functions under the assignment of edge-to-edge gains (as in \cite{algebraic}) are shown in Fig. \ref{fig:bflow}. 
The formulation described in \cite{algebraic} gives the following 8 equations in 10 variables:
\begin{align*}
&\alpha_3+\alpha_4\alpha_1=1 &\alpha_4\alpha_2=0\\
&\alpha_5+\alpha_6\alpha_1=0 &\alpha_6\alpha_2=1\\
&\alpha_7\alpha_2+\alpha_8=0 &\alpha_7\alpha_1=1 \\
&\alpha_9\alpha_2+\alpha_{10}=1 &\alpha_9\alpha_1=0
\end{align*}

In this work, we propose methods to simplify the algebraic formulation for the general scalar-linear network coding problem through the use of path gains as opposed to edge-to-edge gains used in \cite{algebraic}. As shown in the remainder of the paper, the path gain approach results in considerable simplifications in several cases.  
\section{Algebraic Formulation Using Path Gains}
\label{sec:transform}
The main idea in the proposed formulation is to use path gains instead of edge-to-edge gains as variables and obtain a system of polynomial equations. We begin by showing a derivation of the path gain formulation from the edge-to-edge gain formulation.
\subsection{Derivation from Koetter-M\'{e}dard formulation}
\label{sec:deriv-from-koett}
Let the output edge function at the sink $t_j$ demanding source $s(j)$ be $\mathbf{f}_{t_j}=[g(1)\;g(2)\cdots g(|S|)]$. If $P=(e_1e_2\cdots e_l)$ is a path from the source virtual incoming edge $e(s_i)=e_1$ to the sink virtual outgoing edge $e(t_j)=e_l$, the polynomial $g(i)$ contains a path gain term $a(P)=\prod_{m=2}^{l}\alpha_{e_{m-1}e_{m}}$. Conversely, each term in the polynomial $g(i)$ is the gain along a path from the source edge $e(s_i)$ to the sink edge $e(t_j)$.  

In the proposed formulation, the path gain of a path $P$ from a source input virtual edge to a sink output virtual edge is assigned as a variable denoted $a(P)$. Suppose there are $N_{ij}$ paths, denoted $P_{ijk}$ ($1\leq k\leq N_{ij}$), from $e(s_i)$ to $e(t_j)$. We see that the polynomial $g(i)$ can be written as $g(i)=\sum_{k=1}^{N_{ij}}a(P_{ijk})$. 

The proposed approach can be summarized as follows. Equating the output edge function at the sinks to unit vectors, the equations in the Koetter-M\'{e}dard formulation become linear in the new path gain variables. We call these conditions as no-interference conditions. However, if two paths overlap in one or more edges, there are inter-relationships between the path gain variables. These inter-relationships are called edge compatibility conditions, and they turn out to be quadratic in the path gain variables. 

A simple description of the edge compatibility conditions is as follows. If two source-sink paths $P=P_1eP_2$ and $Q=Q_1eQ_2$ overlap in an edge $e$, we see that the relationship $a(P)a(Q)=a(P_1eQ_2)a(Q_1eP_2)$ needs to be satisfied, since both sides are equal to $a(P_1e)a(eP_2)a(Q_1e)a(eQ_2)$. Note that $P_1eQ_2$ and $Q_1eP_2$ are source-sink paths as well. However, several of these equations can be combined to produce the necessary set of edge compatibility conditions. This is described in more detail in the Section \ref{sec:ecc}.

\subsection{Constructing source-sink paths as trees}
To work with the path gain formulation for a given network and connection requirements, we need to determine source-sink paths and assign path gain variables. Then, the no interference conditions and the non-trivial edge compatibility conditions have to be determined. In the remainder of this section, we provide algorithms for performing these tasks. In these algorithms, we employ a graph transformation that is very useful in both visualizing the path gain approach and solving for the edge-to-edge gains from the path gains.

An important ingredient in the algorithms is an ordering of the nodes. Every directed acyclic network determines a topological order or sequencing of nodes from sources or sinks or vice versa. A standard algorithm for finding such a topological ordering of the nodes is given below \cite{deo} for completeness.

\begin{algo}
\label{algo:toposort}
Topological Sorting\\
\textit{Input}: A directed acyclic graph, $G=(V,E)$.
\begin{enumerate}
\item Associate with each node $v$, a value $N(v)$ that is initialized to $|O(v)|$.
\item Pick a node $v$ such that $N(v)=0$, do
\begin{itemize}
\item For each edge $e \in I(v)$,\\ $N(\text{tail}(e)) \leftarrow N(\text{tail}(e)) - 1$. 
\item $N(v) \leftarrow -1$
\item Append $v$ to the ordering, $P$.
\end{itemize}
\item If any node has not been added to the ordering yet, go to Step 2. Else terminate.
\end{enumerate}
\textit{Output}: $P$, a topological ordering of nodes.
\end{algo}
Notice that the sinks occur first in the topological ordering. Loosely, the ordering traverses the nodes from sinks to sources. The final algorithm that takes a network coding problem as input and outputs a set of trees that collect together all source-sink paths is given below:

\begin{algo}
Graph Transformation\\
\noindent \textit{Input}: A directed acyclic graph $G=(V,E)$, set of sources $S$, set of sinks $T$, connection requirements $\mathcal{C}$.
\begin{enumerate}
\item Obtain a topological ordering $P$ for the graph $G=(V,E)$ using Algorithm 1.
\item Let $G'(V',E')=G(V,E)$.
\item Loop through the nodes $v \in V$ in the order defined by $P$, do
\begin{itemize}
\item If $O(v)>1$, 
	\begin{itemize}
	\item for each edge $e \in O(v)$, add a new node $v'$ to $V'$ with one output link connecting it to $\text{head}(e)$ and one input link $e'$ for each $e'' \in I(v)$ such that $\text{tail}(e')=\text{tail}(e'')$.
        \item Delete the old node $v$ in $V'$.
	\end{itemize}
\end{itemize}
\end{enumerate}
\noindent \textit{Output}: $G'=(V',E')$, a transformed network.
\end{algo}

\begin{thm}
The final transformed network is a set of $|T|$ directed trees $\{T_1,T_2,\cdots,T_j\}$ such that Sink $t_j$ is the root of the $j$-th tree. All leaf nodes in the trees are copies of one of the source nodes. There is a one-to-one correspondence between the paths from leaf nodes, which are copies of the source $s_i$, to the root in $T_j$ and the paths from $s_i$ to $t_j$ in the original network.
\end{thm}
\begin{IEEEproof}
Each node in the transformed network will have exactly one output link and the acyclic property of the graph is maintained by the transformation. The underlying undirected graph is a set of disjoint trees, because any cycle in it must imply that either the cycle is also present in the directed graph or that one of the nodes in the directed graph has more than one output link. Hence, the equivalent network is made up of a set of directed trees.

The transformation maintains one output link for each node in the original graph that has $|O(v)| \geq 1$. So, the only nodes that will have $|O(v)|=0$, and hence be the roots of these trees, are the sink nodes (which had $|O(v)|=0$ to start with). Hence, each sink would be the root of a directed tree in which all edges are directed towards this root.

Also, the number of input links of a copied node in the transformed graph is equal to the number of input links possessed by the original node. So, the only nodes that will have $|I(v)|=0$, and hence be leaf nodes in these trees, are copies of the source nodes (which had $|I(v)|=0$ to start with).

Finally, since all nodes are visited in the topological order from the sinks to the sources, all paths from the sinks to the sources will be part of the final network. This results in the one-to-one correspondence in the paths.
\end{IEEEproof}

\begin{figure*}[t]
\centering
\begin{picture}(0,0)%
\includegraphics{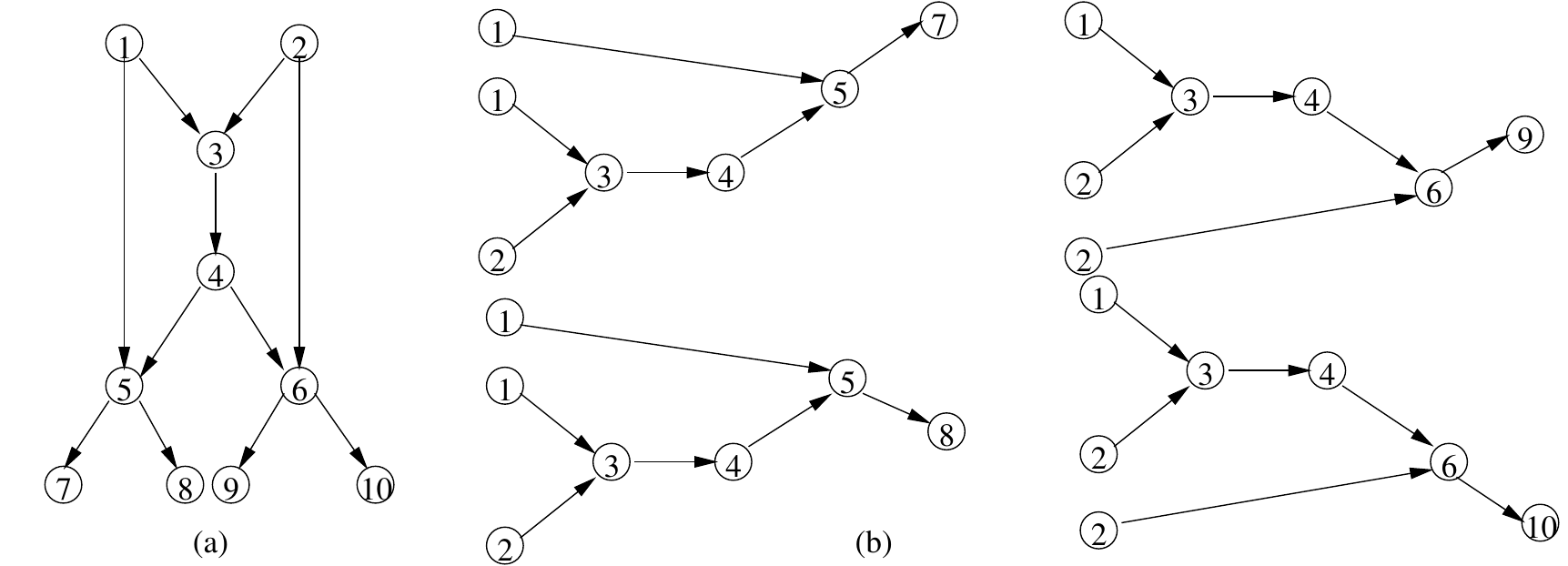}%
\end{picture}%
\setlength{\unitlength}{3522sp}%
\begingroup\makeatletter\ifx\SetFigFont\undefined%
\gdef\SetFigFont#1#2#3#4#5{%
  \reset@font\fontsize{#1}{#2pt}%
  \fontfamily{#3}\fontseries{#4}\fontshape{#5}%
  \selectfont}%
\fi\endgroup%
\begin{picture}(9245,3339)(-599,-4003)
\put(5941,-3166){\makebox(0,0)[lb]{\smash{{\SetFigFont{10}{12.0}{\rmdefault}{\mddefault}{\updefault}{\color[rgb]{0,0,0}$e_2$}%
}}}}
\put(5941,-2671){\makebox(0,0)[lb]{\smash{{\SetFigFont{10}{12.0}{\rmdefault}{\mddefault}{\updefault}{\color[rgb]{0,0,0}$e_1$}%
}}}}
\put(6751,-2761){\makebox(0,0)[lb]{\smash{{\SetFigFont{10}{12.0}{\rmdefault}{\mddefault}{\updefault}{\color[rgb]{0,0,0}$e_3$}%
}}}}
\put(7381,-3256){\makebox(0,0)[lb]{\smash{{\SetFigFont{10}{12.0}{\rmdefault}{\mddefault}{\updefault}{\color[rgb]{0,0,0}$e_7$}%
}}}}
\put(8011,-3706){\makebox(0,0)[lb]{\smash{{\SetFigFont{10}{12.0}{\rmdefault}{\mddefault}{\updefault}{\color[rgb]{0,0,0}$e_{11}$}%
}}}}
\put(6661,-3526){\makebox(0,0)[lb]{\smash{{\SetFigFont{10}{12.0}{\rmdefault}{\mddefault}{\updefault}{\color[rgb]{0,0,0}$e_5$}%
}}}}
\put(5311,-2581){\makebox(0,0)[lb]{\smash{{\SetFigFont{10}{12.0}{\rmdefault}{\mddefault}{\updefault}{\color[rgb]{0,0,0}$a_6X_1$}%
}}}}
\put(5311,-3391){\makebox(0,0)[lb]{\smash{{\SetFigFont{10}{12.0}{\rmdefault}{\mddefault}{\updefault}{\color[rgb]{0,0,0}$b_5X_2$}%
}}}}
\put(5311,-3931){\makebox(0,0)[lb]{\smash{{\SetFigFont{10}{12.0}{\rmdefault}{\mddefault}{\updefault}{\color[rgb]{0,0,0}$b_6X_2$}%
}}}}
\put(2431,-3706){\makebox(0,0)[lb]{\smash{{\SetFigFont{10}{12.0}{\rmdefault}{\mddefault}{\updefault}{\color[rgb]{0,0,0}$e_2$}%
}}}}
\put(2431,-3211){\makebox(0,0)[lb]{\smash{{\SetFigFont{10}{12.0}{\rmdefault}{\mddefault}{\updefault}{\color[rgb]{0,0,0}$e_1$}%
}}}}
\put(3241,-3301){\makebox(0,0)[lb]{\smash{{\SetFigFont{10}{12.0}{\rmdefault}{\mddefault}{\updefault}{\color[rgb]{0,0,0}$e_3$}%
}}}}
\put(3871,-3121){\makebox(0,0)[lb]{\smash{{\SetFigFont{10}{12.0}{\rmdefault}{\mddefault}{\updefault}{\color[rgb]{0,0,0}$e_6$}%
}}}}
\put(3286,-2671){\makebox(0,0)[lb]{\smash{{\SetFigFont{10}{12.0}{\rmdefault}{\mddefault}{\updefault}{\color[rgb]{0,0,0}$e_4$}%
}}}}
\put(1846,-2581){\makebox(0,0)[lb]{\smash{{\SetFigFont{10}{12.0}{\rmdefault}{\mddefault}{\updefault}{\color[rgb]{0,0,0}$a_3X_1$}%
}}}}
\put(1846,-3031){\makebox(0,0)[lb]{\smash{{\SetFigFont{10}{12.0}{\rmdefault}{\mddefault}{\updefault}{\color[rgb]{0,0,0}$a_4X_1$}%
}}}}
\put(1846,-3931){\makebox(0,0)[lb]{\smash{{\SetFigFont{10}{12.0}{\rmdefault}{\mddefault}{\updefault}{\color[rgb]{0,0,0}$b_2X_2$}%
}}}}
\put(4591,-3031){\makebox(0,0)[lb]{\smash{{\SetFigFont{10}{12.0}{\rmdefault}{\mddefault}{\updefault}{\color[rgb]{0,0,0}$e_9$}%
}}}}
\put(271,-1051){\makebox(0,0)[lb]{\smash{{\SetFigFont{10}{12.0}{\rmdefault}{\mddefault}{\updefault}{\color[rgb]{0,0,0}$e_1$}%
}}}}
\put(451,-1906){\makebox(0,0)[lb]{\smash{{\SetFigFont{10}{12.0}{\rmdefault}{\mddefault}{\updefault}{\color[rgb]{0,0,0}$e_3$}%
}}}}
\put(226,-2491){\makebox(0,0)[lb]{\smash{{\SetFigFont{10}{12.0}{\rmdefault}{\mddefault}{\updefault}{\color[rgb]{0,0,0}$e_6$}%
}}}}
\put(856,-2491){\makebox(0,0)[lb]{\smash{{\SetFigFont{10}{12.0}{\rmdefault}{\mddefault}{\updefault}{\color[rgb]{0,0,0}$e_7$}%
}}}}
\put(-314,-3166){\makebox(0,0)[lb]{\smash{{\SetFigFont{10}{12.0}{\rmdefault}{\mddefault}{\updefault}{\color[rgb]{0,0,0}$e_8$}%
}}}}
\put(316,-3166){\makebox(0,0)[lb]{\smash{{\SetFigFont{10}{12.0}{\rmdefault}{\mddefault}{\updefault}{\color[rgb]{0,0,0}$e_9$}%
}}}}
\put(631,-3166){\makebox(0,0)[lb]{\smash{{\SetFigFont{10}{12.0}{\rmdefault}{\mddefault}{\updefault}{\color[rgb]{0,0,0}$e_{10}$}%
}}}}
\put(1396,-3166){\makebox(0,0)[lb]{\smash{{\SetFigFont{10}{12.0}{\rmdefault}{\mddefault}{\updefault}{\color[rgb]{0,0,0}$e_{11}$}%
}}}}
\put(811,-1051){\makebox(0,0)[lb]{\smash{{\SetFigFont{10}{12.0}{\rmdefault}{\mddefault}{\updefault}{\color[rgb]{0,0,0}$e_2$}%
}}}}
\put(-584,-3661){\makebox(0,0)[lb]{\smash{{\SetFigFont{10}{12.0}{\rmdefault}{\mddefault}{\updefault}{\color[rgb]{0,0,0}$X_1$}%
}}}}
\put(136,-3616){\makebox(0,0)[lb]{\smash{{\SetFigFont{10}{12.0}{\rmdefault}{\mddefault}{\updefault}{\color[rgb]{0,0,0}$X_2$}%
}}}}
\put(901,-3616){\makebox(0,0)[lb]{\smash{{\SetFigFont{10}{12.0}{\rmdefault}{\mddefault}{\updefault}{\color[rgb]{0,0,0}$X_1$}%
}}}}
\put(1216,-3616){\makebox(0,0)[lb]{\smash{{\SetFigFont{10}{12.0}{\rmdefault}{\mddefault}{\updefault}{\color[rgb]{0,0,0}$X_2$}%
}}}}
\put(-134,-1726){\makebox(0,0)[lb]{\smash{{\SetFigFont{10}{12.0}{\rmdefault}{\mddefault}{\updefault}{\color[rgb]{0,0,0}$e_4$}%
}}}}
\put(991,-1726){\makebox(0,0)[lb]{\smash{{\SetFigFont{10}{12.0}{\rmdefault}{\mddefault}{\updefault}{\color[rgb]{0,0,0}$e_5$}%
}}}}
\put(5851,-1546){\makebox(0,0)[lb]{\smash{{\SetFigFont{10}{12.0}{\rmdefault}{\mddefault}{\updefault}{\color[rgb]{0,0,0}$e_2$}%
}}}}
\put(5851,-1051){\makebox(0,0)[lb]{\smash{{\SetFigFont{10}{12.0}{\rmdefault}{\mddefault}{\updefault}{\color[rgb]{0,0,0}$e_1$}%
}}}}
\put(6661,-1141){\makebox(0,0)[lb]{\smash{{\SetFigFont{10}{12.0}{\rmdefault}{\mddefault}{\updefault}{\color[rgb]{0,0,0}$e_3$}%
}}}}
\put(7291,-1636){\makebox(0,0)[lb]{\smash{{\SetFigFont{10}{12.0}{\rmdefault}{\mddefault}{\updefault}{\color[rgb]{0,0,0}$e_7$}%
}}}}
\put(7921,-1501){\makebox(0,0)[lb]{\smash{{\SetFigFont{10}{12.0}{\rmdefault}{\mddefault}{\updefault}{\color[rgb]{0,0,0}$e_{10}$}%
}}}}
\put(6571,-1906){\makebox(0,0)[lb]{\smash{{\SetFigFont{10}{12.0}{\rmdefault}{\mddefault}{\updefault}{\color[rgb]{0,0,0}$e_5$}%
}}}}
\put(5311,-871){\makebox(0,0)[lb]{\smash{{\SetFigFont{10}{12.0}{\rmdefault}{\mddefault}{\updefault}{\color[rgb]{0,0,0}$a_5X_1$}%
}}}}
\put(5311,-1771){\makebox(0,0)[lb]{\smash{{\SetFigFont{10}{12.0}{\rmdefault}{\mddefault}{\updefault}{\color[rgb]{0,0,0}$b_3X_2$}%
}}}}
\put(5311,-2176){\makebox(0,0)[lb]{\smash{{\SetFigFont{10}{12.0}{\rmdefault}{\mddefault}{\updefault}{\color[rgb]{0,0,0}$b_4X_2$}%
}}}}
\put(2386,-1996){\makebox(0,0)[lb]{\smash{{\SetFigFont{10}{12.0}{\rmdefault}{\mddefault}{\updefault}{\color[rgb]{0,0,0}$e_2$}%
}}}}
\put(2386,-1501){\makebox(0,0)[lb]{\smash{{\SetFigFont{10}{12.0}{\rmdefault}{\mddefault}{\updefault}{\color[rgb]{0,0,0}$e_1$}%
}}}}
\put(3196,-1591){\makebox(0,0)[lb]{\smash{{\SetFigFont{10}{12.0}{\rmdefault}{\mddefault}{\updefault}{\color[rgb]{0,0,0}$e_3$}%
}}}}
\put(3826,-1411){\makebox(0,0)[lb]{\smash{{\SetFigFont{10}{12.0}{\rmdefault}{\mddefault}{\updefault}{\color[rgb]{0,0,0}$e_6$}%
}}}}
\put(4411,-916){\makebox(0,0)[lb]{\smash{{\SetFigFont{10}{12.0}{\rmdefault}{\mddefault}{\updefault}{\color[rgb]{0,0,0}$e_8$}%
}}}}
\put(3241,-961){\makebox(0,0)[lb]{\smash{{\SetFigFont{10}{12.0}{\rmdefault}{\mddefault}{\updefault}{\color[rgb]{0,0,0}$e_4$}%
}}}}
\put(1801,-871){\makebox(0,0)[lb]{\smash{{\SetFigFont{10}{12.0}{\rmdefault}{\mddefault}{\updefault}{\color[rgb]{0,0,0}$a_1X_1$}%
}}}}
\put(1801,-1321){\makebox(0,0)[lb]{\smash{{\SetFigFont{10}{12.0}{\rmdefault}{\mddefault}{\updefault}{\color[rgb]{0,0,0}$a_2X_1$}%
}}}}
\put(1801,-2221){\makebox(0,0)[lb]{\smash{{\SetFigFont{10}{12.0}{\rmdefault}{\mddefault}{\updefault}{\color[rgb]{0,0,0}$b_1X_2$}%
}}}}
\put(8326,-1276){\makebox(0,0)[lb]{\smash{{\SetFigFont{10}{12.0}{\rmdefault}{\mddefault}{\updefault}{\color[rgb]{0,0,0}$X_1$}%
}}}}
\put(4906,-1186){\makebox(0,0)[lb]{\smash{{\SetFigFont{10}{12.0}{\rmdefault}{\mddefault}{\updefault}{\color[rgb]{0,0,0}$X_1$}%
}}}}
\put(4771,-3526){\makebox(0,0)[lb]{\smash{{\SetFigFont{10}{12.0}{\rmdefault}{\mddefault}{\updefault}{\color[rgb]{0,0,0}$X_2$}%
}}}}
\put(8326,-3526){\makebox(0,0)[lb]{\smash{{\SetFigFont{10}{12.0}{\rmdefault}{\mddefault}{\updefault}{\color[rgb]{0,0,0}$X_2$}%
}}}}
\end{picture}%
\caption{(a) The (modified) butterfly network with 4 sinks and 2 sources. (b) The final transformed network has 4 disjoint trees - one for each sink node. The path gains at the leaf nodes are denoted by $a_{i}$'s and $b_{i}$'s.}
\label{butterfly}
\end{figure*}
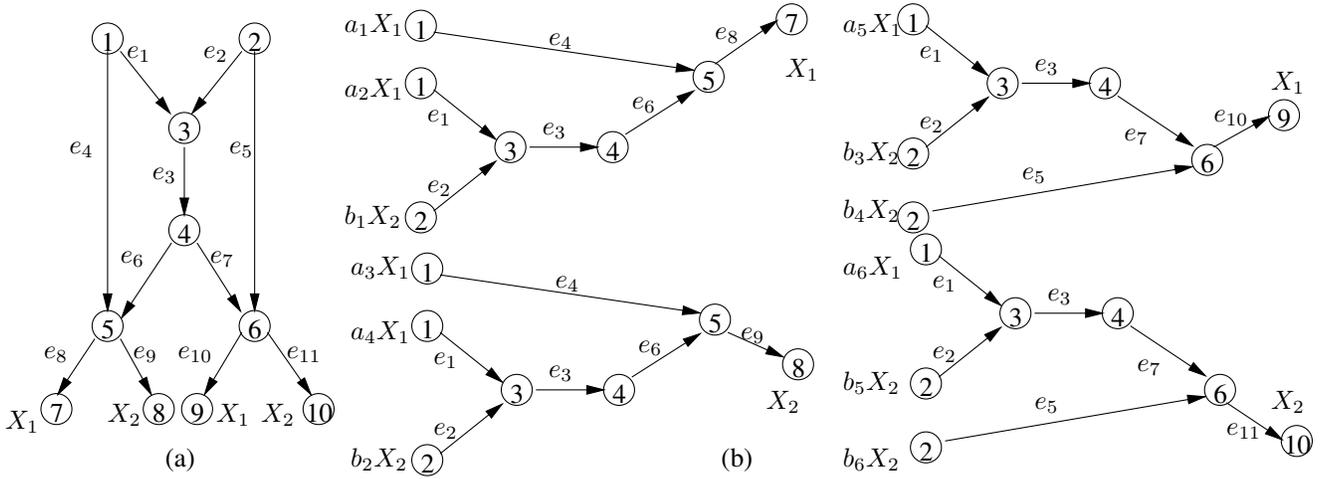

An example of this transformation applied to the butterfly network (Fig.~\ref{butterfly}a) can be seen in Fig.~\ref{butterfly}b. 
To apply the graph transformation, the topological ordering of the nodes is chosen to be $7-8-9-10-5-6-4-3-1-2$. Nodes 7, 8, 9 and 10 are sink nodes, and occur first in the ordering. Nodes 5 and 6 will be replicated 2 times, since they both have 2 output links. This will result in the replication of the edges $e_4$, $e_5$, $e_6$ and $e_7$. Node 4 will now have 4 output links and will have to be replicated as many times along with edge $e_3$. Similarly, Node 3 will also be replicated 4 times along with edges $e_1$ and $e_2$. Finally, the source nodes 1 and 2 will be replicated 6 times each since they both now have 6 output links.

\subsection{Path gain variables and edge functions}
Since there is a one-to-one correspondence between the leaf source nodes in the transformed network and the source-sink paths in the original network, path gain variables are assigned at the leaf source nodes. The assignment is illustrated in Fig.~\ref{butterfly}b for the butterfly network. Source nodes 1 and 2 are assigned the variable names $a$ and $b$, respectively. The subscripts are chosen tree by tree in the transformed network. In the tree with root as Node 7, the two copies of source node 1 are assigned variables $a_1$ and $a_2$, while the single copy of source node 2 is assigned the variable $b_1$. In the tree with root node 8, the variables are $a_3$, $a_4$ for the two copies of Node 1, and $b_2$ for the single copy of Node 2. We continue in this manner to name the scaling variables at the source leaf nodes of the other two trees to get variables $a_1$, $a_2$, $\cdots$, $a_6$ and $b_1$, $b_2$, $\cdots$, $b_6$.  

Once path gain variables are assigned (from some field) at the leaf nodes, all edge functions are computed in the transformed network assuming that intermediate nodes perform addition only. The output function at the root (sink) is the sum of all incoming edge functions. For instance, in the tree with root as Node 7 in Fig. \ref{butterfly}b, the edge functions are as follows: for $e_1$, $a_2X_1$; for $e_2$, $b_1X_2$; for $e_3$ and $e_6$, $a_2X_1+b_1X_2$; for $e_4$, $a_1X_1$; for $e_8$, the edge function is $(a_1+a_2)X_1+b_1X_2$. The output function at sink node 7 is $(a_1+a_2)X_1+b_1X_2$. Similarly, the edge functions can be computed for the other trees. Note that the intermediate nodes perform addition as the entire path gain has been assigned as a variable at the leaf.

\subsection{No Interference conditions}
\label{sec:nic}
Because of the equivalence between paths from sources to a sink $t_j$ in the original network and leaf nodes in the tree $T_j$, we see that the output function calculated in the transformed network is identical to the output function in the Koetter-M\'{e}dard formulation as given in Section \ref{sec:deriv-from-koett}. Therefore, the no interference conditions are obtained by equating the output function of the root in the transformed network to its demand.

In Fig. \ref{butterfly}b, the output function at the root nodes 7, 8, 9 and 10 are $(a_1+a_2)X_1+b_1X_2$, $(a_3+a_4)X_1+b_2X_2$, $a_5X_1+(b_3+b_4)X_2$ and $a_6X_1+(b_5+b_6)X_2$, respectively. For the symbol at Node 7 to be equal to the required $X_1$, we have $a_1+a_2=1$ and $b_1=0$. Other equations are derived similarly. Hence, in the butterfly network of Fig. \ref{butterfly}, we get the following linear equations:

\begin{align}
\label{butter:ni}
&a_1+a_2=1 &b_1=0\nonumber\\
&a_3+a_4=0 &b_2=1\nonumber\\
&a_5=1 &b_3+b_4=0\nonumber\\
&a_6=0 &b_5+b_6=1
\end{align}

For completion, we state the general form of the no-interference conditions below. In general, each path gain variable in the transformed network is associated with exactly one source symbol and one sink (or tree). Let us denote the source-sink path gains by $a_{ijk}$ where $i\in \{1,\ldots,|S|\}$ denotes the source, $j\in \{1,\ldots,|T|\}$ denotes the sink (or the tree), and $k\in {1,\ldots,N_{ij}}$ is an index among all copies of the source node $s_i$ in the tree $T_j$ rooted at $t_j$. Then, the general form of the ``No Interference'' conditions can be written as follows:
\begin{equation}
\sum_ka_{ijk} =\left\{
\begin{array}{lll}
1 & \text{if}\ s(j)=s_i\\
0 & \text{otherwise}\\
\end{array}\right.
\end{equation}

\subsection{Edge Compatibility conditions}
\label{sec:ecc}
As explained before, the path gain variables of overlapping paths are related by quadratic edge-compatibility conditions. If multiple copies of an edge are present in the transformed network, then the edge is part of multiple source-sink paths in the original network. Therefore, edge compatibility conditions are indicated by the presence of multiple edges in the transformed network. The edge functions in the transformed network can be used to write down the edge compatibility conditions. We first show this for the butterfly network example and later provide the general form.

In our illustrative example of Fig.~\ref{butterfly}b, the edge $e_3$ is copied four times. Since there are $\binom{4}{2}=6$ ways of choosing two copies among the four, there will be six edge compatibility conditions for $e_3$. The symbols on the copies of $e_3$ on the trees with root nodes 7, 8, 9 and 10 are $a_2X_1+b_1X_2$, $a_4X_1+b_2X_2$, $a_5X_1+b_3X_2$ and $a_6X_1+b_5X_2$, respectively. Hence, in fractional form, we need $\frac{a_2}{a_4}=\frac{b_1}{b_2}$ (roots 7 and 8), $\frac{a_2}{a_5}=\frac{b_1}{b_3}$ (roots 7 and 9), $\frac{a_2}{a_6}=\frac{b_1}{b_5}$ (roots 7 and 10), $\frac{a_4}{a_5}=\frac{b_2}{b_3}$ (8 and 9), $\frac{a_4}{a_6}=\frac{b_2}{b_5}$ (8 and 10) and $\frac{a_5}{a_6}=\frac{b_3}{b_5}$ (9 and 10).  

In the degree-2 form, the edge compatibility conditions for the four copies of the edge $e_3$ are listed below:
\begin{align}
\label{butter:ec}
&a_2b_2=a_4b_1 & a_2b_3=a_5b_1\nonumber\\
&a_2b_5=a_6b_1 & a_4b_3=a_5b_2\nonumber\\
&a_4b_5=a_6b_2 & a_5b_5=a_6b_3
\end{align}
For the butterfly network example, we do not get any other edge compatibility conditions. For edges $e_6$ and $e_7$, the equations are identical to the ones listed above. Also, there are no equations for edges $e_1$, $e_2$, $e_4$ and $e_5$ since these edges have scaled versions of the same symbol flowing through them. 

We have seen that not all duplicated edges result in distinct compatibility conditions. In general, edge compatibility equations will be required for each edge $e$ in the original network that satisfies the following conditions:
\begin{enumerate}
\item Number of copies of $\text{head}(e)$ in the transformed network $>\ 1$ (or the edge will not be replicated at all).
\item Number of different source nodes having a path to $e\ >\ 1$ (since if two copies of $e$ carry $a_1X_1$ and $a_2X_1$, these will be scalar multiples of each other for any value assigned to $a_1,a_2$).
\item $|I(\text{tail}(e))|>1$ (or the equations will be same as that for $e' \in I(\text{tail}(e))$).
\end{enumerate}

We now state the general form of the edge-compatibility conditions in terms of nodes of the transformed network. Given a node $v \in V$ in the original network, the general form of the condition for two copies of $v$, denoted by $v_1$ and $v_2$ in $V'$, belonging to the $j_1$-th and $j_2$-th trees, respectively, can be written as follows:
\begin{align}
\label{eq:genedge}
&\lefteqn{\left(\sum_{k\in h_{i_1j_1}(v_1)}a_{i_1j_1k}\right) \left(\sum_{l\in h_{i_2j_2}(v_2)}a_{i_2j_2l}\right) =}\nonumber\\
& &\left(\sum_{m\in h_{i_1j_2}(v_2)}a_{i_1j_2m}\right) \left(\sum_{n\in h_{i_2j_1}(v_1)}a_{i_2j_1n}\right)
\end{align}
where $h_{ij}(v)$ denotes the set of leaf nodes in the $j$-th tree that are copies of the source node $s_i$ and have a path to $v$. 

A careful study of the general form shows an edge compatibility condition needs to be introduced for every two copies, $v_1,v_2\in V'$, of node $v\in V$ and for every two sources $s_{i_1}, s_{i_2}\in S$ such that (a) $|I(v_1)|>1$, (b) $v_1\in V_{j_1}, v_2\in V_{j_2},\ V_{j_i}=$ Set of nodes in the $j_i$-th tree, and (c) $h_{i_1j_1}(v_1)\neq \phi, h_{i_2j_1}(v_1)\neq \phi$.

The linear no-interference conditions and the quadratic edge compatibility conditions on the path gains are necessary and sufficient conditions for existence of solutions to the scalar linear network coding problem. The sufficiency is proved by Algorithm \ref{algo:reverse} in Section \ref{sec:reverse}. Before describing the sufficiency, we show how the linear and quadratic equations in path gains can be simplified in a systematic manner to provide useful results.
\subsection{Simplifying the equations}
The linear equations (No Interference conditions) possess the special property that each of them involves a mutually exclusive set of variables. Using this property, we can simplify the system of equations in the following two ways:
\begin{enumerate}
\item It is possible that some of the variables never occur in the non-linear equations (Edge Compatibility conditions). From \eqref{butter:ec}, we can see that $a_1$ is one such variable in the example of the butterfly network. It can be easily seen that the linear equation involving $a_1$ can be trivially satisfied for any value assigned to the other variables involved in the same linear equation by choosing an appropriate value of $a_1$ (which does not have any other condition on it). Hence, $a_1$ along with the linear equation it occurs in can be removed from the system as trivially solvable.\\
Therefore, the first simplification would involve elimination of variables (and their corresponding linear equations) that do not occur in any non-linear equation.
\item Since each linear equation involves a mutually exclusive set of variables, we can eliminate one variable using each linear equation easily. Eliminating this variable from the non-linear equations (note that this does not increase the degree of the system) might reduce some of them to linear equations which can again be used to eliminate more variables iteratively.
\end{enumerate}

In the case of the butterfly network, after the first step of simplification, we are left with 8 variables, 4 linear equations and 6 non-linear equations.

In the second step of the simplification, after the first round of elimination of variables using the linear equations (\ref{butter:ni}) in (\ref{butter:ec}), we are left with 4 variables: $a_2$, $a_4$, $b_3$ and $b_5$ and the 6 equations as shown below. 
\begin{align*}
&a_2=0&b_5=0\\
&a_2b_5=0&a_2b_3=0\\
&a_4b_5=0&a_4b_3=1
\end{align*}
Subsequently, $a_2$ and $b_5$ can also be eliminated, using the linear equations above, leaving just 2 variables and the relation:
\begin{equation}\label{eq:butfinal}
a_4b_3=1
\end{equation}

Hence, the network coding problem for the example of the butterfly network has been reduced to solving only one (non-trivial) equation given in \eqref{eq:butfinal}.

\section{Illustrative Examples}
\label{sec:example}
In this section, we provide a few examples to illustrate the usefulness of the path gain approach in deriving the system of polynomial equations corresponding to a network coding problem. Note that several problems in this area of polynomial equations and network coding are NP-hard or undecidable, and we do not expect polynomial-time algorithms and exact step-by-step solutions to result from the path gain approach. Our approach is to demonstrate the effectiveness of the path gain method in several examples of varying complexity. 

In all examples, we provide the number of equations and variables obtained from the edge-to-edge gain formulation. The path gain formulation (after simplifications) will result in better numbers in many cases. However, we point out that this is not a comparison of the two methods, since one is simplified and the other is not. As we have shown, the path gain method can be seen as a method for simplifying the edge-to-edge gain equations. Other methods for simplifying generic systems of polynomial equations, such as Gr\"{o}bner basis methods, are useful in several networks. Also, Gr\"{o}bner basis or other methods can be used after the path-gain-based simplifications. However, in many examples, we observe that the path gain formulation appears to provide results on solvability. This is mainly because the path gain approach provides low degree equations, which are amenable to easy analysis and further simplifications. 
\subsection{Illustration of derivation from Koetter-M\'{e}dard formulation}
For the butterfly network, the relationship between the path gain variables (shown in Fig. \ref{butterfly}b) and the edge-to-edge gains in the Koetter-M\'{e}dard formulation (Fig. \ref{fig:bflow}) can be written down as follows: $a_1=\alpha_3$ (path: $1-5-7$), $a_2=\alpha_4\alpha_1$ (path: $1-3-5-7$), $b_1=\alpha_4\alpha_2$ (path: $2-3-4-5-7$), $a_3=\alpha_5$ (path: $1-5-8$), $a_4=\alpha_6\alpha_1$ (path: $1-3-4-5-8$), $b_2=\alpha_6\alpha_2$ (path: $2-3-4-5-8$), $a_5=\alpha_1\alpha_7$ (path: $1-3-4-6-9$), $b_3=\alpha_2\alpha_7$ (path: $2-3-4-6-9$), $b_4=\alpha_8$ (path: $2-6-9$), $a_6=\alpha_1\alpha_9$ (path: $1-3-4-6-10$), $b_5=\alpha_2\alpha_9$ (path: $2-3-4-6-10$), $b_6=\alpha_{10}$ (path: $2-6-10$). 
\begin{figure*}
\centering
\begin{picture}(0,0)%
\includegraphics{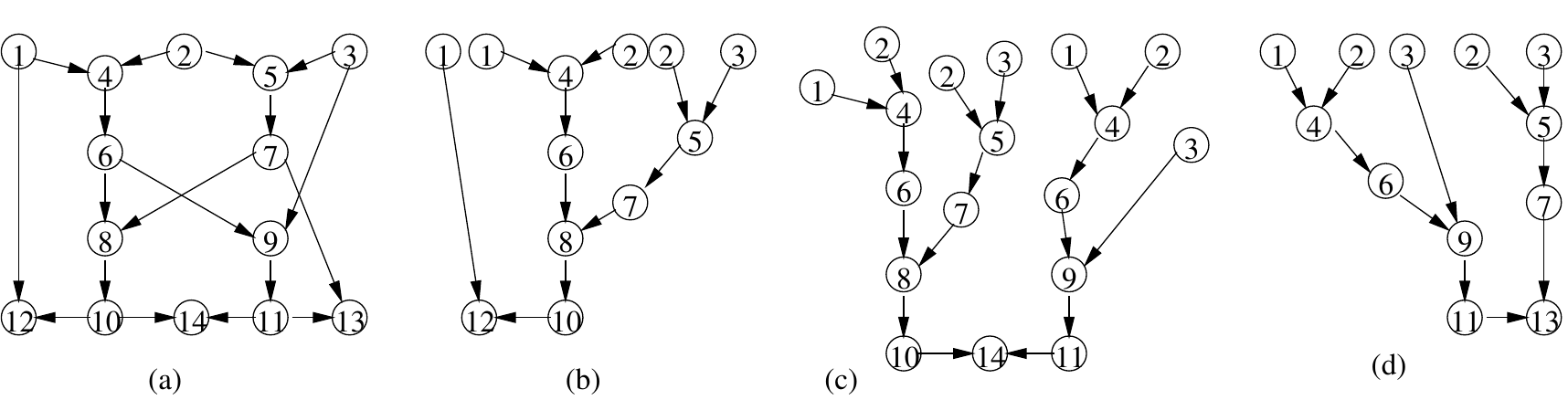}%
\end{picture}%
\setlength{\unitlength}{3315sp}%
\begingroup\makeatletter\ifx\SetFigFont\undefined%
\gdef\SetFigFont#1#2#3#4#5{%
  \reset@font\fontsize{#1}{#2pt}%
  \fontfamily{#3}\fontseries{#4}\fontshape{#5}%
  \selectfont}%
\fi\endgroup%
\begin{picture}(9797,2494)(19,-3095)
\put(451,-1321){\makebox(0,0)[lb]{\smash{{\SetFigFont{10}{12.0}{\rmdefault}{\mddefault}{\updefault}{\color[rgb]{0,0,0}$e_1$}%
}}}}
\put(451,-2311){\makebox(0,0)[lb]{\smash{{\SetFigFont{10}{12.0}{\rmdefault}{\mddefault}{\updefault}{\color[rgb]{0,0,0}$e_2$}%
}}}}
\put(1486,-2311){\makebox(0,0)[lb]{\smash{{\SetFigFont{10}{12.0}{\rmdefault}{\mddefault}{\updefault}{\color[rgb]{0,0,0}$e_3$}%
}}}}
\put(1486,-1276){\makebox(0,0)[lb]{\smash{{\SetFigFont{10}{12.0}{\rmdefault}{\mddefault}{\updefault}{\color[rgb]{0,0,0}$e_4$}%
}}}}
\put(3331,-1321){\makebox(0,0)[lb]{\smash{{\SetFigFont{10}{12.0}{\rmdefault}{\mddefault}{\updefault}{\color[rgb]{0,0,0}$e_1$}%
}}}}
\put(3331,-2311){\makebox(0,0)[lb]{\smash{{\SetFigFont{10}{12.0}{\rmdefault}{\mddefault}{\updefault}{\color[rgb]{0,0,0}$e_2$}%
}}}}
\put(4006,-1546){\makebox(0,0)[lb]{\smash{{\SetFigFont{10}{12.0}{\rmdefault}{\mddefault}{\updefault}{\color[rgb]{0,0,0}$e_4$}%
}}}}
\put(5446,-1546){\makebox(0,0)[lb]{\smash{{\SetFigFont{10}{12.0}{\rmdefault}{\mddefault}{\updefault}{\color[rgb]{0,0,0}$e_1$}%
}}}}
\put(5446,-2536){\makebox(0,0)[lb]{\smash{{\SetFigFont{10}{12.0}{\rmdefault}{\mddefault}{\updefault}{\color[rgb]{0,0,0}$e_2$}%
}}}}
\put(6481,-2536){\makebox(0,0)[lb]{\smash{{\SetFigFont{10}{12.0}{\rmdefault}{\mddefault}{\updefault}{\color[rgb]{0,0,0}$e_3$}%
}}}}
\put(5941,-1591){\makebox(0,0)[lb]{\smash{{\SetFigFont{10}{12.0}{\rmdefault}{\mddefault}{\updefault}{\color[rgb]{0,0,0}$e_4$}%
}}}}
\put(6661,-1456){\makebox(0,0)[lb]{\smash{{\SetFigFont{10}{12.0}{\rmdefault}{\mddefault}{\updefault}{\color[rgb]{0,0,0}$e_1$}%
}}}}
\put(8956,-2311){\makebox(0,0)[lb]{\smash{{\SetFigFont{10}{12.0}{\rmdefault}{\mddefault}{\updefault}{\color[rgb]{0,0,0}$e_3$}%
}}}}
\put(9451,-1591){\makebox(0,0)[lb]{\smash{{\SetFigFont{10}{12.0}{\rmdefault}{\mddefault}{\updefault}{\color[rgb]{0,0,0}$e_4$}%
}}}}
\put(8281,-1636){\makebox(0,0)[lb]{\smash{{\SetFigFont{10}{12.0}{\rmdefault}{\mddefault}{\updefault}{\color[rgb]{0,0,0}$e_1$}%
}}}}
\put(4951,-1006){\makebox(0,0)[lb]{\smash{{\SetFigFont{8}{9.6}{\familydefault}{\mddefault}{\updefault}{\color[rgb]{0,0,0}$a_3$}%
}}}}
\put(5356,-736){\makebox(0,0)[lb]{\smash{{\SetFigFont{8}{9.6}{\familydefault}{\mddefault}{\updefault}{\color[rgb]{0,0,0}$b_3$}%
}}}}
\put(7426,-1321){\makebox(0,0)[lb]{\smash{{\SetFigFont{8}{9.6}{\familydefault}{\mddefault}{\updefault}{\color[rgb]{0,0,0}$c_3$}%
}}}}
\put(7876,-736){\makebox(0,0)[lb]{\smash{{\SetFigFont{8}{9.6}{\familydefault}{\mddefault}{\updefault}{\color[rgb]{0,0,0}$a_5$}%
}}}}
\put(8326,-736){\makebox(0,0)[lb]{\smash{{\SetFigFont{8}{9.6}{\familydefault}{\mddefault}{\updefault}{\color[rgb]{0,0,0}$b_6$}%
}}}}
\put(9091,-736){\makebox(0,0)[lb]{\smash{{\SetFigFont{8}{9.6}{\familydefault}{\mddefault}{\updefault}{\color[rgb]{0,0,0}$b_7$}%
}}}}
\put(8731,-736){\makebox(0,0)[lb]{\smash{{\SetFigFont{8}{9.6}{\familydefault}{\mddefault}{\updefault}{\color[rgb]{0,0,0}$c_4$}%
}}}}
\put(9586,-736){\makebox(0,0)[lb]{\smash{{\SetFigFont{8}{9.6}{\familydefault}{\mddefault}{\updefault}{\color[rgb]{0,0,0}$c_5$}%
}}}}
\put(2971,-736){\makebox(0,0)[lb]{\smash{{\SetFigFont{8}{9.6}{\familydefault}{\mddefault}{\updefault}{\color[rgb]{0,0,0}$a_2$}%
}}}}
\put(2566,-736){\makebox(0,0)[lb]{\smash{{\SetFigFont{8}{9.6}{\familydefault}{\mddefault}{\updefault}{\color[rgb]{0,0,0}$a_1$}%
}}}}
\put(4141,-736){\makebox(0,0)[lb]{\smash{{\SetFigFont{8}{9.6}{\familydefault}{\mddefault}{\updefault}{\color[rgb]{0,0,0}$b_2$}%
}}}}
\put(4591,-736){\makebox(0,0)[lb]{\smash{{\SetFigFont{8}{9.6}{\familydefault}{\mddefault}{\updefault}{\color[rgb]{0,0,0}$c_1$}%
}}}}
\put(3826,-736){\makebox(0,0)[lb]{\smash{{\SetFigFont{8}{9.6}{\familydefault}{\mddefault}{\updefault}{\color[rgb]{0,0,0}$b_1$}%
}}}}
\put(5806,-871){\makebox(0,0)[lb]{\smash{{\SetFigFont{8}{9.6}{\familydefault}{\mddefault}{\updefault}{\color[rgb]{0,0,0}$b_4$}%
}}}}
\put(6211,-781){\makebox(0,0)[lb]{\smash{{\SetFigFont{8}{9.6}{\familydefault}{\mddefault}{\updefault}{\color[rgb]{0,0,0}$c_2$}%
}}}}
\put(6571,-736){\makebox(0,0)[lb]{\smash{{\SetFigFont{8}{9.6}{\familydefault}{\mddefault}{\updefault}{\color[rgb]{0,0,0}$a_4$}%
}}}}
\put(7156,-736){\makebox(0,0)[lb]{\smash{{\SetFigFont{8}{9.6}{\familydefault}{\mddefault}{\updefault}{\color[rgb]{0,0,0}$b_5$}%
}}}}
\end{picture}%
\caption{(a) An example network that is solvable only over fields with characteristic 2. There are three sources - 1, 2 and 3 - producing symbols $X_1$, $X_2$ and $X_3$ respectively. There are three sinks - 12, 13 and 14 - demanding symbols $X_3$, $X_1$ and $X_2$ respectively. (b),(c),(d) The final transformed network with 3 trees - one for each sink node.}
\label{fig:char2}
\end{figure*}
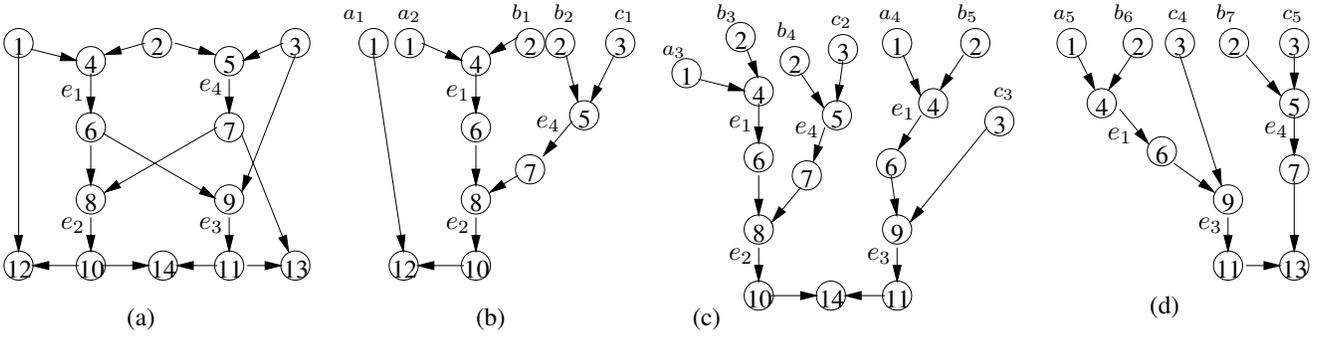

The no-interference conditions are easily obtained. For edge compatibility between the paths $2-3-4-5-8$ and the path $1-3-4-6-9$, we get the equation $b_2a_5=b_3a_4=\alpha_6\alpha_2\alpha_1\alpha_7$. Other compatibility conditions can be checked similarly.

The change to path gain variables results in easy simplification of the resulting equations with no increase in degree. Finally, we obtain the simple equation, $a_4b_3=1$, which is not obvious even when the substitution is clearly specified.   
\subsection{Another Example}
Consider the network shown in Fig.~\ref{fig:char2}a taken from \cite{doughinsuff,doughpoly}, where it has been proved to have linear coding solutions only over fields of characteristic 2. 
Nodes 1, 2 and 3 are sources producing $X_1$, $X_2$ and $X_3$ respectively. Nodes 12, 13 and 14 are sinks demanding $X_3$, $X_1$ and $X_2$ respectively. The trees in the equivalent transformed network are shown in Fig.~\ref{fig:char2}b,c,d. 

The set of equations generated by the ``No Interference condition'' are:
\begin{align}
&\mbox{Node 12: }a_1+a_2=0; b_1+b_2=0; c_1=1\nonumber\\
&\mbox{Node 13: }a_3=1; b_3+b_4=0; c_2+c_3=0\nonumber\\
&\mbox{Node 14: }a_4+a_5=0; b_5+b_6+b_7=1; c_4+c_5=0
\end{align}

The set of equations generated by the ``Edge Compatibility condition'' for edges $e_1$, $e_2$, $e_3$ and $e_4$ respectively are:
\begin{align}
e_1:\ &a_2b_3=a_3b_1; a_2b_5=a_4b_1; a_2b_7=a_5b_1;\nonumber\\
&a_3b_5=a_4b_3; a_3b_7=a_5b_3; a_4b_7=a_5b_5\nonumber\\
e_2:\ &a_2(b_5+b_6)=a_4(b_1+b_2); a_2c_4=a_4c_1;\nonumber\\
&(b_1+b_2)c_4=(b_5+b_6)c_1\nonumber\\
e_3:\ &a_3b_7=a_5b_3; a_3c_5=a_5c_2; b_3c_5=b_7c_2\nonumber\\
e_4:\ &b_2c_3=b_4c_1; b_2c_4=b_6c_1; b_4c_4=b_6c_3
\end{align}

Using the linear equations to eliminate variables iteratively, we get 9 equations in 6 variables shown below.
\begin{IEEEeqnarray}{l}
a_2b_3=b_1;\ a_2=-a_4b_1;\ a_4b_3=-1;\IEEEnonumber\\
a_2c_4=a_4;\ c_4=a_4c_2;\ b_3c_4+c_2=0;\IEEEnonumber\\
b_1c_2+b_3=0;\ b_1c_4+1=0;\ b_3c_4=c_2\IEEEyesnumber
\end{IEEEeqnarray}

From equations $b_3c_4+c_2=0$ and $b_3c_4=c_2$, we can derive the relation $2c_2=0$. Substituting $c_2=0$ in the above system leads to the condition $1=0$, which is not possible. Hence, we must have $2=0$, which implies that the system is not solvable in any field with an odd characteristic. Also, in characteristic 2, setting all variables to 1 in the above equations, is seen to be a solution. This example demonstrates that, in practice, working with the equations derived through the path gain formulation can be advantageous. 

For this example, the Koetter-M\'{e}dard formulation, as illustrated in \cite{doughpoly}, results in 17 equations in 22 variables. However, as shown in \cite{doughpoly}, it is possible to derive $2=0$ from these 17 equations using other simplifications. Alternatively, a Gr\"{o}bner basis method can also be used to derive $2=0$. The path gain approach should be seen as a generic technique for simplification that can be used in arbitrary network coding problems, as shown in the next two examples.
\subsection{Multicast example}
\begin{figure}[htb]
  \centering
\begin{picture}(0,0)%
\includegraphics{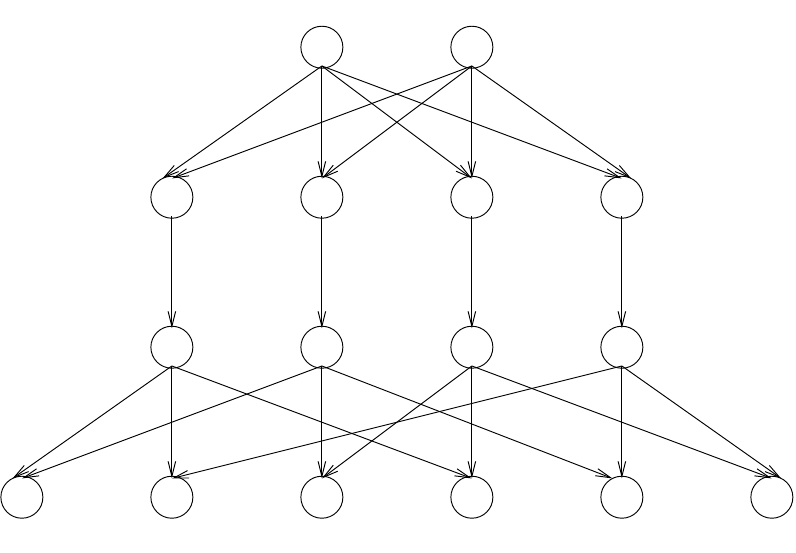}%
\end{picture}%
\setlength{\unitlength}{2368sp}%
\begingroup\makeatletter\ifx\SetFigFont\undefined%
\gdef\SetFigFont#1#2#3#4#5{%
  \reset@font\fontsize{#1}{#2pt}%
  \fontfamily{#3}\fontseries{#4}\fontshape{#5}%
  \selectfont}%
\fi\endgroup%
\begin{picture}(6352,4411)(425,-4400)
\put(6526,-4036){\makebox(0,0)[lb]{\smash{{\SetFigFont{7}{8.4}{\familydefault}{\mddefault}{\updefault}{\color[rgb]{0,0,0}16}%
}}}}
\put(3001,-436){\makebox(0,0)[b]{\smash{{\SetFigFont{7}{8.4}{\familydefault}{\mddefault}{\updefault}{\color[rgb]{0,0,0}1}%
}}}}
\put(4201,-436){\makebox(0,0)[b]{\smash{{\SetFigFont{7}{8.4}{\familydefault}{\mddefault}{\updefault}{\color[rgb]{0,0,0}2}%
}}}}
\put(3001,-1636){\makebox(0,0)[b]{\smash{{\SetFigFont{7}{8.4}{\familydefault}{\mddefault}{\updefault}{\color[rgb]{0,0,0}4}%
}}}}
\put(4201,-1636){\makebox(0,0)[b]{\smash{{\SetFigFont{7}{8.4}{\familydefault}{\mddefault}{\updefault}{\color[rgb]{0,0,0}5}%
}}}}
\put(5401,-1636){\makebox(0,0)[b]{\smash{{\SetFigFont{7}{8.4}{\familydefault}{\mddefault}{\updefault}{\color[rgb]{0,0,0}6}%
}}}}
\put(5401,-2836){\makebox(0,0)[b]{\smash{{\SetFigFont{7}{8.4}{\familydefault}{\mddefault}{\updefault}{\color[rgb]{0,0,0}10}%
}}}}
\put(4201,-2836){\makebox(0,0)[b]{\smash{{\SetFigFont{7}{8.4}{\familydefault}{\mddefault}{\updefault}{\color[rgb]{0,0,0}9}%
}}}}
\put(3001,-2836){\makebox(0,0)[b]{\smash{{\SetFigFont{7}{8.4}{\familydefault}{\mddefault}{\updefault}{\color[rgb]{0,0,0}8}%
}}}}
\put(3001,-4036){\makebox(0,0)[b]{\smash{{\SetFigFont{7}{8.4}{\familydefault}{\mddefault}{\updefault}{\color[rgb]{0,0,0}13}%
}}}}
\put(4201,-4036){\makebox(0,0)[b]{\smash{{\SetFigFont{7}{8.4}{\familydefault}{\mddefault}{\updefault}{\color[rgb]{0,0,0}14}%
}}}}
\put(5401,-4036){\makebox(0,0)[b]{\smash{{\SetFigFont{7}{8.4}{\familydefault}{\mddefault}{\updefault}{\color[rgb]{0,0,0}15}%
}}}}
\put(1801,-1636){\makebox(0,0)[b]{\smash{{\SetFigFont{7}{8.4}{\familydefault}{\mddefault}{\updefault}{\color[rgb]{0,0,0}3}%
}}}}
\put(1801,-2836){\makebox(0,0)[b]{\smash{{\SetFigFont{7}{8.4}{\familydefault}{\mddefault}{\updefault}{\color[rgb]{0,0,0}7}%
}}}}
\put(1801,-4036){\makebox(0,0)[b]{\smash{{\SetFigFont{7}{8.4}{\familydefault}{\mddefault}{\updefault}{\color[rgb]{0,0,0}12}%
}}}}
\put(601,-4036){\makebox(0,0)[b]{\smash{{\SetFigFont{7}{8.4}{\familydefault}{\mddefault}{\updefault}{\color[rgb]{0,0,0}11}%
}}}}
\put(3001,-136){\makebox(0,0)[b]{\smash{{\SetFigFont{7}{8.4}{\familydefault}{\mddefault}{\updefault}{\color[rgb]{0,0,0}$X_1$}%
}}}}
\put(4201,-136){\makebox(0,0)[b]{\smash{{\SetFigFont{7}{8.4}{\familydefault}{\mddefault}{\updefault}{\color[rgb]{0,0,0}$X_2$}%
}}}}
\put(601,-4336){\makebox(0,0)[b]{\smash{{\SetFigFont{7}{8.4}{\familydefault}{\mddefault}{\updefault}{\color[rgb]{0,0,0}$X_1, X_2$}%
}}}}
\put(1801,-4336){\makebox(0,0)[b]{\smash{{\SetFigFont{7}{8.4}{\familydefault}{\mddefault}{\updefault}{\color[rgb]{0,0,0}$X_1, X_2$}%
}}}}
\put(3001,-4336){\makebox(0,0)[b]{\smash{{\SetFigFont{7}{8.4}{\familydefault}{\mddefault}{\updefault}{\color[rgb]{0,0,0}$X_1, X_2$}%
}}}}
\put(4201,-4336){\makebox(0,0)[b]{\smash{{\SetFigFont{7}{8.4}{\familydefault}{\mddefault}{\updefault}{\color[rgb]{0,0,0}$X_1, X_2$}%
}}}}
\put(5401,-4336){\makebox(0,0)[b]{\smash{{\SetFigFont{7}{8.4}{\familydefault}{\mddefault}{\updefault}{\color[rgb]{0,0,0}$X_1, X_2$}%
}}}}
\put(6601,-4336){\makebox(0,0)[b]{\smash{{\SetFigFont{7}{8.4}{\familydefault}{\mddefault}{\updefault}{\color[rgb]{0,0,0}$X_1, X_2$}%
}}}}
\end{picture}%
  \caption{Multicast example.}
  \label{fig:multex}
\end{figure}
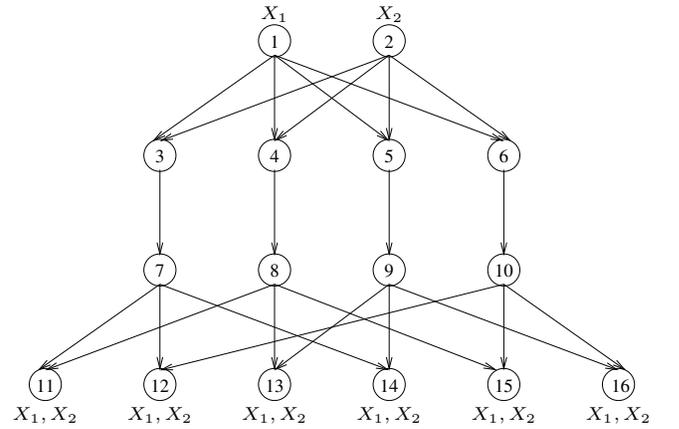

An interesting example of a multicast problem, presented in \cite{Dougherty:2004cr}, is shown in Fig. \ref{fig:multex}.
The sources are nodes 1 and 2, and the sinks are nodes 11-16. This problem does not have a binary solution, as shown in \cite{Dougherty:2004cr}.

Using the edge-to-edge gain formulation, we get 24 equations in 32 variables. The path gain method initially results in 84 equations in 48 variables. After the simplifications, we obtain 54 equations in 18 variables. Significantly, there are 6 quadratic equations, each of the form $x^2_3+x_3+x_1x_2=0$. Next, we can show that $x_1x_2=0$ (either $x_1=0$ or $x_2=0$) provides a contradiction in the equations. Hence, we have equations of the form $x^2+x+1=0$, which cannot be solved in the binary field. With some more analysis, we can find solutions over GF(4).

From this example, we see that the path gain formulation provides useful simplifications in non-trivial cases. In contrast, Gr\"{o}bner basis methods on the edge-to-edge gain equations are not immediately useful in showing linear in-solvability over GF(2). Note that this does not rule out any other simplification of the edge-to-edge equations to obtain the necessary result. We merely conclude that the path gain method provides a useful simplification.
\subsection{A Bigger Example}
\label{sec:bigger-example}
\begin{figure*}
\centering
\includegraphics[width=5in]{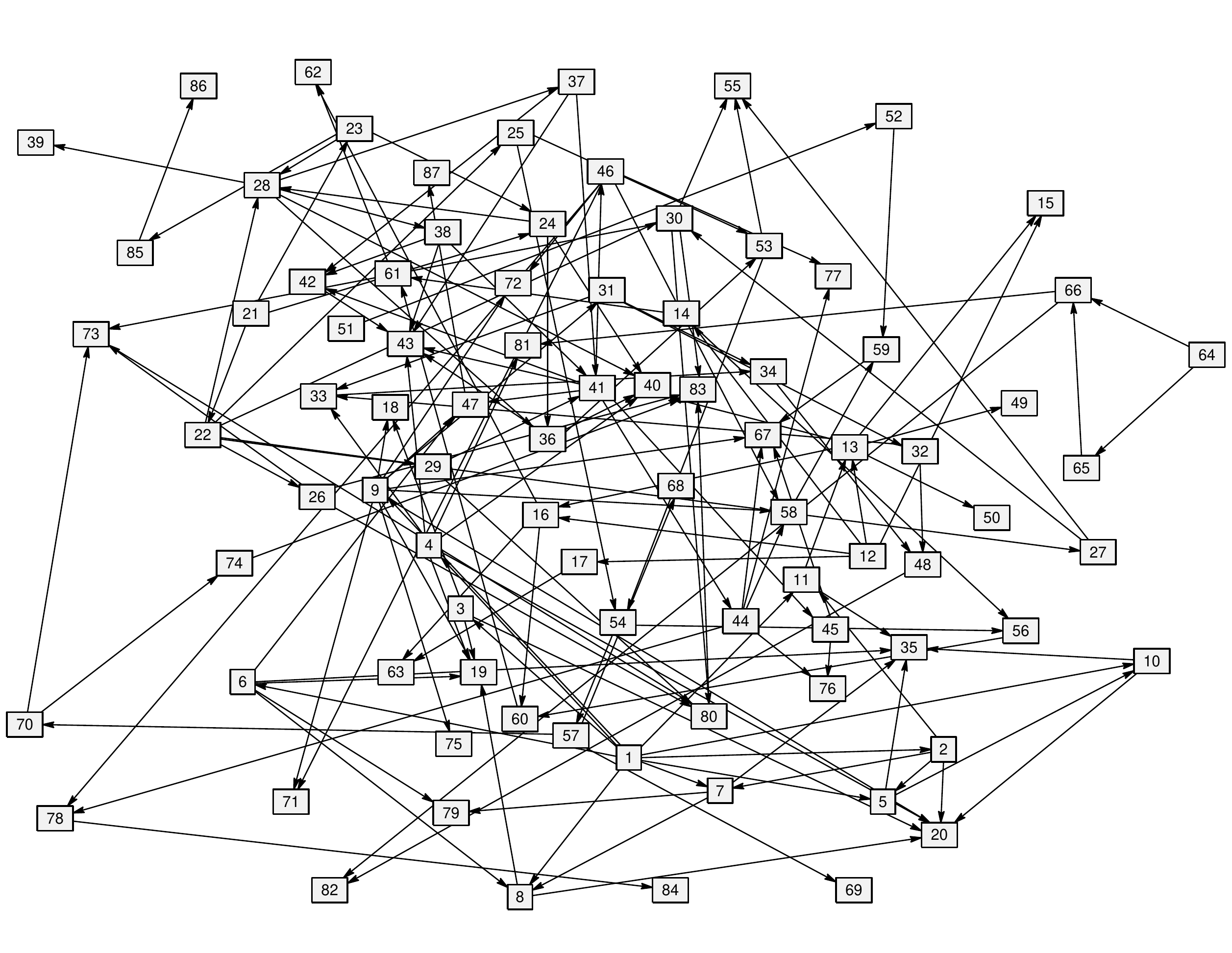}
\caption{An ISP network over Europe with 87 nodes and 161 edges.}
\label{isp}
\end{figure*}

Consider an ISP network topology shown in Fig.~\ref{isp} taken from \cite{rocketfuel}. 

The network has 87 nodes and 161 edges. Edges are directed from lower-numbered nodes to higher-numbered nodes i.e. in an edge $(u,v)$, $u<v$. Hence, the graph is directed and acyclic. We assume all links have unit capacity, and use fields of characteristic 2 in our simplification steps. After directing the graph, the five nodes 1, 12, 21, 51 and 64 were set as sources in the example problems. Sink nodes and demands were chosen at random from among the sources visible from each sink. The graph is not reduced by this choice of demands, since all nodes are visible from the five chosen sources. 
\begin{enumerate}
\item {\it 5 sources (all rate 1), 10 sinks (all rate 1).} The edge-to-edge gain formulation gives a system of 44 equations in 30 variables. The path gain formulation initially results in 44 linear equations and 3 degree-2 equations in 316 variables. After applying the simplification steps, we are left with only 3 degree-2 equations in 7 variables assuming solution exists in a characteristic 2 field. In fact, setting all the remaining 7 variables to zero results in a valid solution to the three equations (some other scaling variables are non-zero). Hence, a solution over GF(2) is possible. 
\item {\it 5 sources (one with rate 2, others rate 1), 11 sinks (all rate 1).} The edge-to-edge gain formulation yields a system of 50 equations in 40 variables in this case. In comparison, the path gain formulation initially resulted in 50 linear equations and 34 degree-2 equations in 330 variables. But after applying the simplification steps, we are left with only 13 degree-2 equations in 17 variables assuming solution exists in a characteristic 2 field. Again the all-zero solution is valid for the remaining 17 variables resulting in a network code over GF(2).
\item {\it 5 sources (all with rate 2), 8 sinks (all rate 1).} The edge-to-edge gain formulation yields a system of 88 equations in 180 variables. The path gain formulation initially gives 88 linear and 11198 degree-2 equations in 632 variables. But on applying the simplification steps, assuming characteristic 2, it turns out that the system is not solvable over characteristic 2.
\end{enumerate}

To run further examples, we computed the max-flows from the sources (1,12,21,51,64) to a set of 11 nodes chosen as sinks. Rates below the individual max-flows were assigned to sources and sink demands. The results are summarized below (Notation: $S$, $T$ are source and sink sets. A source node $s$ of rate $R>1$ is shown as $R$ source nodes $s_1,\;s_2\;\cdots\;s_R$. The demands of each sink are shown in brackets).
\begin{enumerate}
\item $S=\{1_1,1_2,12_1,12_2, 21, 51, 64\}$, $T=\{15 (12_1,12_2),$ $40 (1_1,12_1,21), 43(21), 49(1_1), 62 (1_1,12_1,21),63 (12_1,$ $12_2), 67(1_1), 71(21), 82(64), 83(21), 86(21)\}.$ 
The path gain formulation yields 1188 equations in  507 variables. After simplifications, there are 476 equations, but none of them have a constant term. Hence, setting the remaining variables to zero provides a binary solution. 
\item $S=\{1_1,1_2,12_1,12_2,21,51,64\}$, $T=\{15 (12_1,12_2),$ $40 (1_1,1_2,12_1,21), 43(21), 49(1_1), 62 (1_1,12_1,12_2,21),$ $63 (12_1,12_2), 67(1_1), 71(21), 82(64), 83(21), 86(21)\}$. We obtain 555 variables and 1683 equations. Upon simplification, we find that a solution does not exist in characteristic 2.
\item $S=\{1,12,21_1,21_2,51,64\}$, $T=\{15(1),40(21_1,$ $21_2),43(21_1), 49(1), 63(12), 67(1), 71(21_1),82(64),$ $83(21_1), 86(21_1)\}$. The path gain formulation yields 578 variables and 12048 equations. After simplifications, there are 6780 equations, but only five of them have a constant term. A linear term (one path gain) appears in each of these five equations, but does not appear as a linear term in any other equation. Hence, setting this one path gain to 1 and the remaining variables to zero provides a binary solution. 
\end{enumerate}
As expected, solutions are not guaranteed even if all demands are within individual max flows. We see that the number of equations and variables increases steeply in some cases. However, guessing a binary solution may be feasible.

To obtain another example, we modified the graph of Fig. \ref{isp} (by changing edge connections) to get a butterfly network as a subgraph when the nodes 62 and 63 demand the sources 1 and 12. On the modified graph, we set $S=\{1,12_1,12_2, 21, 51, 64\}$, $T=\{15$ $(12_1,12_2), 40 (1,12_1,21), 43(21), 49(1), , 62 (1,12_1,21), 63 (1,$ $12_1), 67(1), 71(21), 82(64), 83(21), 86(21)\}$, where the other demands are chosen to be below the individual max flows. We obtain 1247 equations in 503 variables. After simplifications, there are no equations with a constant term. So, setting the remaining variables to zero results in a binary solution. 

Hence, we see that the path-based formulation of scalar linear network coding appears to yield useful results even over large networks with a few sources and sinks. This shows the extent of simplification possible in polynomial systems defined by network coding problems.
\subsection{Simplification summary}
It has been shown in \cite{mollerbound} that the complexity of Gr\"{o}bner Basis algorithms depends, among other things, on the maximum degree of the starting basis. The degrees of the intermediate polynomials computed during Gr\"{o}bner Basis calculations has been shown to grow up to $2^{2^d}$ if the maximum degree of the starting basis is $d$. Due to these issues, Gr\"{o}bner Basis algorithms become practically intractable except for small problem instances. In the light of these results, the path gain formulation that produces degree-2 equations becomes important in reducing the running complexity of Gr\"{o}bner Basis algorithms that may be used to solve the network coding problem.

The simplification provided by the path gain approach is summarized in Table \ref{comparison} for the various examples presented so far. 
\begin{table}[htb]
\caption{Comparison of formulations.}
\label{comparison}
\centering
\begin{tabular}[t]{c|c|c}
\hline
& \it{Path gain} & \it{Edge gains (unsimplified)}\\
{\begin{tabular}[t]{c}
Example\\
\hline\\
Butterfly\\Fig. \ref{fig:char2}a\\\cite[Fig. 5]{algebraic}\\\cite[Fig. 3]{doughinsuff}\\\cite[Fig. 3]{doughpoly}
\end{tabular}
}
&
{\begin{tabular}[t]{ccc}
Var.$^1$ & Deg. 2 Eqns$^1$\\
\hline\\
4 & 6\\8 & 15\\9 & 5\\27 & 45\\12 & 30
\end{tabular}
}
&
{\begin{tabular}[t]{ccc}
Var. & Eqns & Deg.\\
\hline\\
10 & 8 & 2\\14 & 9 & 3\\14 & 4 & 4\\50 & 32 & 3\\22 & 17 & 3
\end{tabular}
}\\
\hline
\end{tabular}
\begin{flushleft}
$^1$After one iteration of elimination using the linear equations
\end{flushleft}
\end{table}
As indicated, we have shown numbers for only one round of linear equation simplification. It can be seen that, apart from having a maximum degree of only 2, the number of variables is also lesser in many cases enabling use of methods such as \cite{xl} for solving the system. 

The number of variables and equations from the edge-to-edge gain formulation are of the order of the number of edges in the network. However, the number of monomial terms possible using the edge-to-edge gain variables is exponential in the number of edges. In a large network, depending on the number of paths from sources to sinks, a large number of monomial terms occur in the system of polynomial equations. Because of the large number of variables and larger number of terms, there is no obvious method to simplify the equations other than running standard routines for Gr\"{o}bner basis. The path gain approach is beneficial in providing results on solvability in some examples and in reducing the complexity of Gr\"{o}bner basis methods in most cases.
\section{Network Code from Path Gain Variables}
\label{sec:reverse}
While the path gain variables are useful for solving the system of polynomial equations, the implementation of network coding is through edge-to-edge gains. We now describe an algorithm to obtain a network code for the original network from the path gain variables in the transformed network. Note that this completes the proof of the sufficiency of edge compatibility conditions. 

First, we will briefly describe the algorithm and then present a notational version of the same. A solution to the system of polynomial equations in the path gain formulation consists of a set of values assigned to the path gain variables at the leaf source nodes in the transformed network such that the no interference conditions as well as the edge compatibility conditions are satisfied. The algorithm to construct a network code from such a solution consists of propagating the values of these coefficients from the source nodes to the sink nodes through the transformed network. 

We compute two vectors for every edge $e$ of the graph $G=(V,E)$. The first vector  $\mathbf{f}_e=[f_e(1)\;f_e(2)\;\cdots\;f_e(|S|)]$ represents the edge function or symbol $\sum_{j=1}^{|S|}f_e(j)X_j$ sent over edge $e$. Suppose $e$ is replicated $n$ times to obtain edges $e''_i$, $1\leq i\leq n$ in the transformed graph $G'=(V',E')$. The second vector $\mathbf{c}_e=[c_1\;c_2\;\cdots\;c_n]$ is such that the edge function on $e''_i\in E'$ is $c_i\sum_{j=1}^{|S|}f_e(j)X_j$. Note that such a scaling property is guaranteed for all copies of an edge by the compatibility conditions. Once the vectors $\mathbf{f}_e$ are computed for all $e\in E$, the network code in $G$ is completely known.

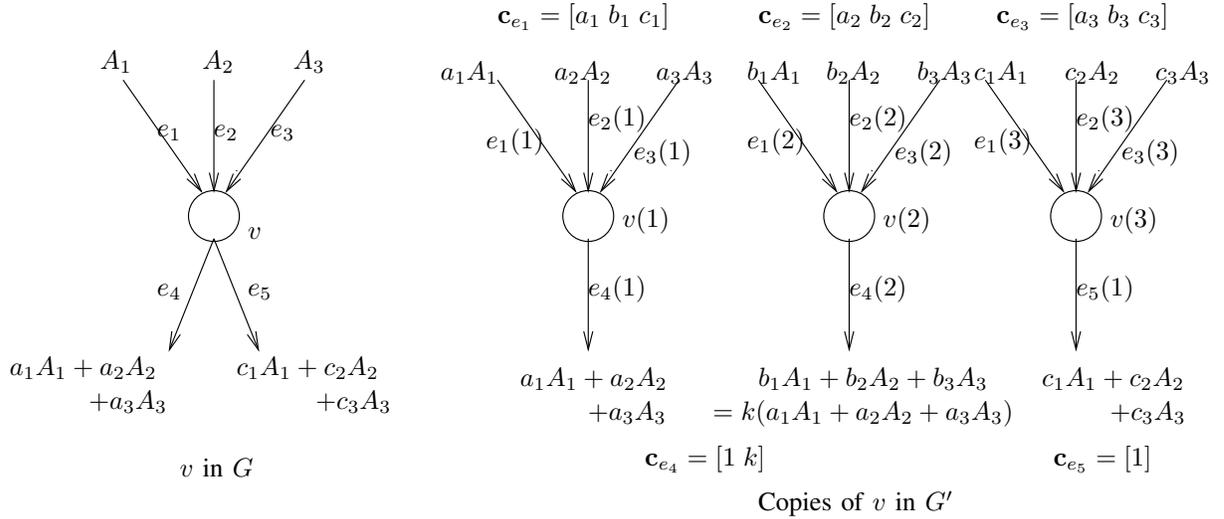
\begin{figure*}[htb]
  \centering
\begin{picture}(0,0)%
\includegraphics{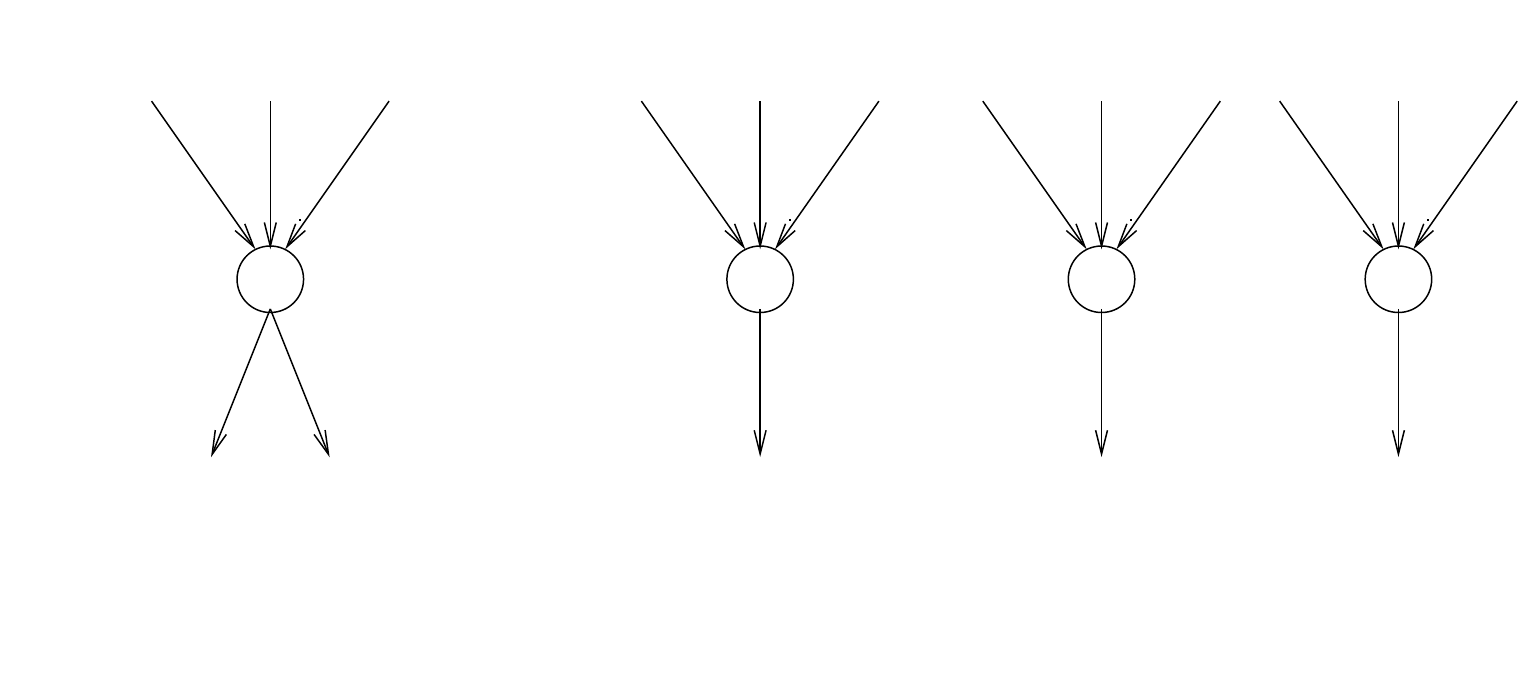}%
\end{picture}%
\setlength{\unitlength}{3750sp}%
\begingroup\makeatletter\ifx\SetFigFont\undefined%
\gdef\SetFigFont#1#2#3#4#5{%
  \reset@font\fontsize{#1}{#2pt}%
  \fontfamily{#3}\fontseries{#4}\fontshape{#5}%
  \selectfont}%
\fi\endgroup%
\begin{picture}(7677,3420)(586,-3196)
\put(1201,-211){\makebox(0,0)[lb]{\smash{{\SetFigFont{10}{12.0}{\familydefault}{\mddefault}{\updefault}{\color[rgb]{0,0,0}$A_1$}%
}}}}
\put(1876,-211){\makebox(0,0)[lb]{\smash{{\SetFigFont{10}{12.0}{\familydefault}{\mddefault}{\updefault}{\color[rgb]{0,0,0}$A_2$}%
}}}}
\put(2476,-211){\makebox(0,0)[lb]{\smash{{\SetFigFont{10}{12.0}{\familydefault}{\mddefault}{\updefault}{\color[rgb]{0,0,0}$A_3$}%
}}}}
\put(1576,-661){\makebox(0,0)[lb]{\smash{{\SetFigFont{10}{12.0}{\familydefault}{\mddefault}{\updefault}{\color[rgb]{0,0,0}$e_1$}%
}}}}
\put(1951,-661){\makebox(0,0)[lb]{\smash{{\SetFigFont{10}{12.0}{\familydefault}{\mddefault}{\updefault}{\color[rgb]{0,0,0}$e_2$}%
}}}}
\put(2326,-661){\makebox(0,0)[lb]{\smash{{\SetFigFont{10}{12.0}{\familydefault}{\mddefault}{\updefault}{\color[rgb]{0,0,0}$e_3$}%
}}}}
\put(2176,-1336){\makebox(0,0)[lb]{\smash{{\SetFigFont{10}{12.0}{\familydefault}{\mddefault}{\updefault}{\color[rgb]{0,0,0}$v$}%
}}}}
\put(2176,-1711){\makebox(0,0)[lb]{\smash{{\SetFigFont{10}{12.0}{\familydefault}{\mddefault}{\updefault}{\color[rgb]{0,0,0}$e_5$}%
}}}}
\put(1576,-1711){\makebox(0,0)[lb]{\smash{{\SetFigFont{10}{12.0}{\familydefault}{\mddefault}{\updefault}{\color[rgb]{0,0,0}$e_4$}%
}}}}
\put(601,-2236){\makebox(0,0)[lb]{\smash{{\SetFigFont{10}{12.0}{\familydefault}{\mddefault}{\updefault}{\color[rgb]{0,0,0}$a_1A_1+a_2A_2$}%
}}}}
\put(1126,-2461){\makebox(0,0)[lb]{\smash{{\SetFigFont{10}{12.0}{\familydefault}{\mddefault}{\updefault}{\color[rgb]{0,0,0}$+a_3A_3$}%
}}}}
\put(2101,-2236){\makebox(0,0)[lb]{\smash{{\SetFigFont{10}{12.0}{\familydefault}{\mddefault}{\updefault}{\color[rgb]{0,0,0}$c_1A_1+c_2A_2$}%
}}}}
\put(2626,-2461){\makebox(0,0)[lb]{\smash{{\SetFigFont{10}{12.0}{\familydefault}{\mddefault}{\updefault}{\color[rgb]{0,0,0}$+c_3A_3$}%
}}}}
\put(5551,-2311){\makebox(0,0)[lb]{\smash{{\SetFigFont{10}{12.0}{\familydefault}{\mddefault}{\updefault}{\color[rgb]{0,0,0}$b_1A_1+b_2A_2+b_3A_3$}%
}}}}
\put(3976,-2311){\makebox(0,0)[lb]{\smash{{\SetFigFont{10}{12.0}{\familydefault}{\mddefault}{\updefault}{\color[rgb]{0,0,0}$a_1A_1+a_2A_2$}%
}}}}
\put(4726,-811){\makebox(0,0)[lb]{\smash{{\SetFigFont{10}{12.0}{\familydefault}{\mddefault}{\updefault}{\color[rgb]{0,0,0}$e_3(1)$}%
}}}}
\put(4426,-586){\makebox(0,0)[lb]{\smash{{\SetFigFont{10}{12.0}{\familydefault}{\mddefault}{\updefault}{\color[rgb]{0,0,0}$e_2(1)$}%
}}}}
\put(3451,-286){\makebox(0,0)[lb]{\smash{{\SetFigFont{10}{12.0}{\familydefault}{\mddefault}{\updefault}{\color[rgb]{0,0,0}$a_1A_1$}%
}}}}
\put(6976,-736){\makebox(0,0)[lb]{\smash{{\SetFigFont{10}{12.0}{\familydefault}{\mddefault}{\updefault}{\color[rgb]{0,0,0}$e_1(3)$}%
}}}}
\put(5476,-736){\makebox(0,0)[lb]{\smash{{\SetFigFont{10}{12.0}{\familydefault}{\mddefault}{\updefault}{\color[rgb]{0,0,0}$e_1(2)$}%
}}}}
\put(4201,-286){\makebox(0,0)[lb]{\smash{{\SetFigFont{10}{12.0}{\familydefault}{\mddefault}{\updefault}{\color[rgb]{0,0,0}$a_2A_2$}%
}}}}
\put(4876,-286){\makebox(0,0)[lb]{\smash{{\SetFigFont{10}{12.0}{\familydefault}{\mddefault}{\updefault}{\color[rgb]{0,0,0}$a_3A_3$}%
}}}}
\put(5476,-286){\makebox(0,0)[lb]{\smash{{\SetFigFont{10}{12.0}{\familydefault}{\mddefault}{\updefault}{\color[rgb]{0,0,0}$b_1A_1$}%
}}}}
\put(6001,-286){\makebox(0,0)[lb]{\smash{{\SetFigFont{10}{12.0}{\familydefault}{\mddefault}{\updefault}{\color[rgb]{0,0,0}$b_2A_2$}%
}}}}
\put(6601,-286){\makebox(0,0)[lb]{\smash{{\SetFigFont{10}{12.0}{\familydefault}{\mddefault}{\updefault}{\color[rgb]{0,0,0}$b_3A_3$}%
}}}}
\put(6151,-586){\makebox(0,0)[lb]{\smash{{\SetFigFont{10}{12.0}{\familydefault}{\mddefault}{\updefault}{\color[rgb]{0,0,0}$e_2(2)$}%
}}}}
\put(6451,-811){\makebox(0,0)[lb]{\smash{{\SetFigFont{10}{12.0}{\familydefault}{\mddefault}{\updefault}{\color[rgb]{0,0,0}$e_3(2)$}%
}}}}
\put(7651,-586){\makebox(0,0)[lb]{\smash{{\SetFigFont{10}{12.0}{\familydefault}{\mddefault}{\updefault}{\color[rgb]{0,0,0}$e_2(3)$}%
}}}}
\put(7951,-811){\makebox(0,0)[lb]{\smash{{\SetFigFont{10}{12.0}{\familydefault}{\mddefault}{\updefault}{\color[rgb]{0,0,0}$e_3(3)$}%
}}}}
\put(7876,-1261){\makebox(0,0)[lb]{\smash{{\SetFigFont{10}{12.0}{\familydefault}{\mddefault}{\updefault}{\color[rgb]{0,0,0}$v(3)$}%
}}}}
\put(6376,-1261){\makebox(0,0)[lb]{\smash{{\SetFigFont{10}{12.0}{\familydefault}{\mddefault}{\updefault}{\color[rgb]{0,0,0}$v(2)$}%
}}}}
\put(4651,-1261){\makebox(0,0)[lb]{\smash{{\SetFigFont{10}{12.0}{\familydefault}{\mddefault}{\updefault}{\color[rgb]{0,0,0}$v(1)$}%
}}}}
\put(4426,-1711){\makebox(0,0)[lb]{\smash{{\SetFigFont{10}{12.0}{\familydefault}{\mddefault}{\updefault}{\color[rgb]{0,0,0}$e_4(1)$}%
}}}}
\put(6151,-1711){\makebox(0,0)[lb]{\smash{{\SetFigFont{10}{12.0}{\familydefault}{\mddefault}{\updefault}{\color[rgb]{0,0,0}$e_4(2)$}%
}}}}
\put(7651,-1711){\makebox(0,0)[lb]{\smash{{\SetFigFont{10}{12.0}{\familydefault}{\mddefault}{\updefault}{\color[rgb]{0,0,0}$e_5(1)$}%
}}}}
\put(1726,-2911){\makebox(0,0)[lb]{\smash{{\SetFigFont{10}{12.0}{\familydefault}{\mddefault}{\updefault}{\color[rgb]{0,0,0}$v$ in $G$}%
}}}}
\put(5551,-3136){\makebox(0,0)[lb]{\smash{{\SetFigFont{10}{12.0}{\familydefault}{\mddefault}{\updefault}{\color[rgb]{0,0,0}Copies of $v$ in $G'$}%
}}}}
\put(3826, 89){\makebox(0,0)[lb]{\smash{{\SetFigFont{10}{12.0}{\familydefault}{\mddefault}{\updefault}{\color[rgb]{0,0,0}$\mathbf{c}_{e_1}=[a_1\;b_1\;c_1]$}%
}}}}
\put(7426,-2311){\makebox(0,0)[lb]{\smash{{\SetFigFont{10}{12.0}{\familydefault}{\mddefault}{\updefault}{\color[rgb]{0,0,0}$c_1A_1+c_2A_2$}%
}}}}
\put(7876,-2536){\makebox(0,0)[lb]{\smash{{\SetFigFont{10}{12.0}{\familydefault}{\mddefault}{\updefault}{\color[rgb]{0,0,0}$+c_3A_3$}%
}}}}
\put(6976,-286){\makebox(0,0)[lb]{\smash{{\SetFigFont{10}{12.0}{\familydefault}{\mddefault}{\updefault}{\color[rgb]{0,0,0}$c_1A_1$}%
}}}}
\put(7576,-286){\makebox(0,0)[lb]{\smash{{\SetFigFont{10}{12.0}{\familydefault}{\mddefault}{\updefault}{\color[rgb]{0,0,0}$c_2A_2$}%
}}}}
\put(8176,-286){\makebox(0,0)[lb]{\smash{{\SetFigFont{10}{12.0}{\familydefault}{\mddefault}{\updefault}{\color[rgb]{0,0,0}$c_3A_3$}%
}}}}
\put(3751,-736){\makebox(0,0)[lb]{\smash{{\SetFigFont{10}{12.0}{\familydefault}{\mddefault}{\updefault}{\color[rgb]{0,0,0}$e_1(1)$}%
}}}}
\put(4426,-2536){\makebox(0,0)[lb]{\smash{{\SetFigFont{10}{12.0}{\familydefault}{\mddefault}{\updefault}{\color[rgb]{0,0,0}$+a_3A_3$}%
}}}}
\put(5251,-2536){\makebox(0,0)[lb]{\smash{{\SetFigFont{10}{12.0}{\familydefault}{\mddefault}{\updefault}{\color[rgb]{0,0,0}$=k(a_1A_1+a_2A_2+a_3A_3)$}%
}}}}
\put(5551, 89){\makebox(0,0)[lb]{\smash{{\SetFigFont{10}{12.0}{\familydefault}{\mddefault}{\updefault}{\color[rgb]{0,0,0}$\mathbf{c}_{e_2}=[a_2\;b_2\;c_2]$}%
}}}}
\put(4801,-2836){\makebox(0,0)[lb]{\smash{{\SetFigFont{10}{12.0}{\familydefault}{\mddefault}{\updefault}{\color[rgb]{0,0,0}$\mathbf{c}_{e_4}=[1\;k]$}%
}}}}
\put(7501,-2836){\makebox(0,0)[lb]{\smash{{\SetFigFont{10}{12.0}{\familydefault}{\mddefault}{\updefault}{\color[rgb]{0,0,0}$\mathbf{c}_{e_5}=[1]$}%
}}}}
\put(7126, 89){\makebox(0,0)[lb]{\smash{{\SetFigFont{10}{12.0}{\familydefault}{\mddefault}{\updefault}{\color[rgb]{0,0,0}$\mathbf{c}_{e_3}=[a_3\;b_3\;c_3]$}%
}}}}
\end{picture}%
  \caption{Determining the vectors $\mathbf{f}$ and $\mathbf{c}$ for outgoing links.}
  \label{fig:reverse}
\end{figure*}

Suppose $\mathbf{f}_{e'}$ and $\mathbf{c}_{e'}$ are known for all the incoming edges $e'\in I(v)$ for a node $v\in V$. The vectors $\mathbf{f}_e$ and $\mathbf{c}_e$ can be computed for the outgoing edges $e\in O(v)$ as illustrated for a sample case in Fig. \ref{fig:reverse}. 
In the figure, a node $v\in V$ with $I(v)=\{e_1,e_2,e_3\}$ and $O(v)=\{e_4,e_5\}$ is replicated thrice into $v(1)$, $v(2)$ and $v(3)$ in $G'$. The incoming and outgoing links are assumed to be replicated as shown in the transformed network. For instance, the edge $e_1$ is replicated thrice as $e_1(1)$, $e_1(2)$ and $e_1(3)$. We suppose that there are three source nodes $S=\{s_1,s_2,s_3\}$, and $\mathbf{f}_{e_i}=[\alpha_{i1}\;\alpha_{i2}\;\alpha_{i3}]$. This results in edge functions $A_i=\sum_{j=1}^3\alpha_{ij}X_j$ for $i=1,2,3$. We assume that the scaling vectors $\mathbf{c}_{e_i}$ are as shown in the figure. 

Using the edge functions and scaling factors on the incoming edges, the edge function of the copies of $e_i,\;i=1,2,3$ are computed first. For instance, the edge function of $e_2(2)$ is computed as $b_2A_2$. Then, the edge function for the outgoing links of $v(1)$, $v(2)$ and $v(3)$ in $G'$ are computed by simple addition. As shown in the figure, the symbols sent on $e_4(1)$ and $e_4(2)$ will be scalar multiples. We then assign the symbol on $e_4$ in $G$ to be the symbol on $e_4(1)$ given by $\sum_{i=1}^3a_iA_i=\sum_{j=1}^3(\sum_{i=1}^3a_i\alpha_{ij})X_j$ (assumed nonzero). Then, $\mathbf{f}_{e_4}$ and $\mathbf{c}_{e_4}$ are assigned suitably. 

In this manner, all the nodes are processed in a suitable order to compute the network code for the original graph from the path gains on the transformed graph. We now introduce some notation to describe the algorithm formally.

\subsection{Notation}
Consider the given network $G=(V,E)$ and the equivalent transformed network $G'=(V',E')$. Then, for each node $v\in V$, let us define the set of network coding coefficients as $a_{e',e}\ \forall\ e'\in I(v), e\in O(v)$ i.e. if $x_{e'}$ is the symbol received on the link $e'\in I(v)$, the symbol sent on $e\in O(v)$ is $\sum_{e'\in I(v)}a_{e',e}x_{e'}$.

Nodes and edges get replicated during the transformation from $G$ to $G'$. We define some sets to hold information about the replicated nodes and edges. For $v\in V$ ($v\notin S\cup T$) and $e,e'\in E$, define:
\begin{align*}
&R_v=\{v'\in V': v' \text{ is a copy of } v\}\\
&R_e=\{e''\in E': e'' \text{ is a copy of } e\}\\
&R_{e',e} =\{e''\in R_{e'}: \exists\; e'''\in R_e\text{ so that head}(e'')=\text{tail}(e''')\}
\end{align*}
The sets $R_v$ and $R_e$ hold nodes and edges in $G'$ that are copies of $v$ and $e$ , respectively. Two such useful sets are (1) $R_{\text{head}(e)}$ that contains copies of $\text{head}(e)$, and (2) $R_{\text{tail}(e)}$ that contains copies of $\text{tail}(e)$. The set $R_{e',e}$ contains copies of an edge $e'$ that connect to a copy of $e$. Clearly, $R_{e',e}$ is non-empty only when $e'\in I(v)$ and $e\in O(v)$ for some node $v$.

For each $e\in I(v)$ in the original graph $G$, there is a one-to-one correspondence between $R_e=\{e_1,e_2,\cdots,e_{|R_e|}\}$ and $R_v=\{v_1,v_2,\cdots,v_{|R_v|}\}$ given by $v_i=\text{head}(e_i)$ in the transformed graph $G'$. Thus, we have the equality $|R_e|=|R_{\text{head}(e)}|$. This is because the incoming edges are duplicated everytime a node is duplicated. So, for $e,e'\in I(v)$ (two incoming edges of one node), we will have $|R_e|=|R_{e'}|$ and the sets $R_e=\{e_1,e_2,\cdots,e_{|R_e|}\}$ and $R_{e'}=\{e'_1,e'_2,\cdots,e'_{|R_{e'}|}\}$ will be ordered such that $v_i=\text{head}(e_i)=\text{head}(e'_i)$.

For each $e\in O(v)$, define the set $R_{v,e}=\{\text{head}(e'):e'\in R_e\}$ to be the subset of $R_v$ that contains nodes whose outgoing edge is a copy of $e$ in $G'$. Note that $|R_{v,e}|=|R_{e}|$.

Let the vector $\mathbf{f}_e=[f_e(1)\;f_e(2)\;\cdots\;f_e(|S|)]$ represent the edge function $\sum_{j=1}^{|S|}f_e(j)X_j$ sent over edge $e\in E$ in the final linear network code in $G$. Since the edge compatibility conditions are satisfied, the edge function on each copy of $e$ in $R_e$ will be a scalar multiple of $\mathbf{f}_e$. For $e''\in R_e$, let the edge function on $e''$ be $\mathbf{f}_{e''}=c_e(e'')\mathbf{f}_e$. We collect the multiplying factors $c_e(e'')$, $e''\in R_e$ into a vector $\mathbf{c}_e=[c_e(e''): e''\in R_e]$. The correspondence between $R_e$ and $R_{\text{head}(e)}$ results in a one-to-one correspondence between elements of the sets $R_{\text{head}(e)}$ and $\mathbf{c}_e$ given by $c_e(e'')\leftrightarrow \text{head}(e'')$ for $e''\in R_e$. 

We define sub-vectors $\mathbf{c}_{e',e}=[c_{e'}(e''):e''\in R_{e',e}]$ collecting the multiplying factors on copies of $e'$ that connect to copies of $e$. For a fixed $e\in O(v)$ and $e',e''\in I(v)$ with $R_{v,e}=\{v_1,v_2,\cdots,v_{|R_{v,e}|}\}$, the sets $R_{e',e}=\{e'_1,e'_2,\cdots,e'_{|R_{e',e}|}\}$ and $R_{e'',e}=\{e''_1,e''_2,\cdots,e''_{|R_{e'',e}|}\}$ will be in one-to-one correspondence and ordered so that $\text{head}(e'_i)=\text{head}(e''_i)=v_i$. So, we have $|R_{e',e}|=|R_{e'',e}|=|R_{v,e}|=|R_e|$. 

In Fig. \ref{fig:reverse}, for instance, we have $R_{e_1}=\{e_1(1),e_1(2),e_1(3)\}$, $R_{e_4}=\{e_4(1),e_4(2)\}$ and $R_{e_5}=\{e_5(1)\}$. Also, $R_{e_1,e_4}=\{e_1(1),e_1(2)\}$ and $R_{e_1,e_5}=\{e_1(3)\}$. Similarily, $R_{e_2,e_4}=\{e_2(1),e_2(2)\}$ and $R_{e_2,e_5}=\{e_2(3)\}$. The scaling vector $\mathbf{c}_{e_1}=[a_1\;b_1\;c_1]$ with $\mathbf{c}_{e_1,e_4}=[a_1\;b_1]$ and $\mathbf{c}_{e_1,e_5}=[c_1]$. Similarily, $\mathbf{c}_{e_2}=[a_2\;b_2\;c_2]$ with $\mathbf{c}_{e_2,e_4}=[a_2\;b_2]$ and $\mathbf{c}_{e_2,e_5}=[c_2]$. Note that all one-to-one correspondences are being preserved in the ordering of coordinates in the scaling vectors.

\noindent{\it Flow matrices at a vertex:} For a vertex $v$, incoming edge $e'\in I(v)$ and outgoing edge $e\in O(v)$, a rank-one flow matrix $F_{e',e}$ is defined as $F_{e',e}= \mathbf{c}_{e',e}^T \mathbf{f}_{e'}$. The matrix $F_{e',e}$ is of dimension $|R_{e',e}|\times|S|$, and the $(i,j)$-th element $F_{e',e}(i,j)=c_{e'}(e''_i)f_{e'}(j)$ (letting $R_{e',e}=\{e''_1,e''_2,\cdots,e''_{|R_{e',e}|}\}$) is the coding coefficient of the $j$-th source symbol flowing in the $i$-th copy of edge $e'$ in $R_{e',e}$. We readily see that each row of $F_{e',e}$ is the coding vector in a copy of edge $e'$ in $G'$. 

In Fig. \ref{fig:reverse}, for instance, we have 
$F_{e_1,e_4}=\begin{bmatrix}a_1\\b_1\end{bmatrix}[A_1]=\begin{bmatrix}a_1\\b_1\end{bmatrix}[\alpha_{11}\;\alpha_{12}\;\alpha_{13}]$ and $F_{e_1,e_5}=[c_1][\alpha_{11}\;\alpha_{12}\;\alpha_{13}]$. In terms of path gain variables, $F_{e',e}(i,j)$ is equal to the sum of the path gain variables for all paths starting from (some copy of) the $j$-th source and using the $i$-th copy of edge $e'$ in $R_{e',e}$. 

Let $I(v)=\{e_1,e_2,\cdots,e_d\}$. For $e\in O(v)$, let $|R_{e_k,e}|=|R_e|=D$ (for all $k$), and let $R_{e_k,e}=\{e'_{k1},e'_{k2},\cdots,e'_{kD}\}$ with $\text{head}(e'_{kl})=v'_l\in R_{v,e}$ independent of $k$. The $l$-th row of the flow matrix $F_{e_k,e}$ contains the flow in the edge $e'_{kl}$ incident on the node $v'_l$ for $1\leq k\leq |S|$. Therefore, the sum $F_e=\sum^d_{k=1}F_{e_k,e}$ is a $D\times |S|$ matrix whose $l$-th row is equal to the sum of all incoming flows into node $v'_l$. By flow conservation, the outgoing flow on the single outgoing edge from node $v'_l$ is equal to the $l$-th row of $F_e$ for $1\leq l\leq D=|R_e|$. So, the rows of $F_e$ contain the flows in the $D$ copies of the edge $e$ in $G'$, and the edge compatibility condition ensures that the matrix $F_e$ is a rank-one matrix. 
\subsection{The Algorithm}
The vectors $\mathbf{f}_e$ and $\mathbf{c}_e$ are initialized for an outgoing link $e$ from the source node as follows. For the $i$-th source node $s_i\in S$ and $e\in O(s_i)$, $\mathbf{f}_e=[0^{i-1}\ 1\ 0^{|S|-i}]$. For $e''\in R_e$, the coordinate $c_e(e'')$ of $\mathbf{c}_e$ is equal to the value of the scaling variable at the source leaf node $\text{tail}(e'')\in R_{s_i}$.
\begin{algo}
\label{algo:reverse}
Deriving the Network Code\\
\textit{Input}: A directed acyclic network $G=(V,E)$, an equivalent transformed network $G'=(V',E')$, a topological ordering of nodes $P$ (from Algorithm \ref{algo:toposort}) and a solution to the derived system of polynomial equations.

\noindent For each node, $v$ in the reverse topological ordering, $P'$, of $P$, if $v\notin S \cup T$, do
\begin{enumerate}
\item Get $\mathbf{f}_{e'},\mathbf{c}_{e'}$ from $\text{tail}(e')\ \forall\ e'\in I(v)$.
\item For each edge $e\in O(v)$
\begin{enumerate}
\item Get $\mathbf{c}_{e',e}$ from $\mathbf{c}_{e'}$ as defined above $\forall\ e'\in I(v)$.
\item $F_{e',e} \leftarrow \mathbf{c}_{e',e}^T \mathbf{f}_{e'}\ \forall\ e'\in I(v)$. 
\item $F_e \leftarrow \sum_{e'\in I(v)}F_{e',e}$, is a matrix such that each row corresponds to the symbol flowing through a copy of edge $e$ in $G'$.
\item $\mathbf{f}_e \leftarrow$ any non-zero row (say, $i$) of $F_e$, or the zero row if $F_e$ is the zero matrix. This is the symbol that will actually flow through $e$ in $G$.
\item $a_{e',e} \leftarrow \mathbf{c}_{e',e}(i)\ \forall\ e'\in I(v)$, where $i$ is the row selected in the previous step. This is the set of network coding coefficients of node $v$ corresponding to output link $e$.
\item $\mathbf{c}_e(j) \leftarrow (j^{th} \mbox{ row of } F_e)/\mathbf{f}_e$ or $0$ if $\mathbf{f}_e=\mathbf{0}\ \forall\ j=1,\ldots,|\mathbf{c}_e|$.
\end{enumerate}
\end{enumerate}
\noindent The decoding coefficients at a sink node $t_j$ are given by the set $\{\mathbf{c}_e; e\in I(t_j)\}$. Note that all the matrices in this set have only one element since there is only one copy of each sink node (and and all its input links) in $G'$.\\
\noindent \textit{Output}: The set of all network coding coefficients, $a_{e',e}$, for the given network.
\end{algo}

In the above algorithm, nodes are travered in the reverse of the topological order obtained from Algorithm \ref{algo:toposort}. At a node $v$, the vectors $\mathbf{f}_{e}$ and $\mathbf{c}_e$ are computed for $e\in O(v)$ using $\mathbf{f}_{e'}$ and $\mathbf{c}_{e'}$ for $e'\in I(v)$. The reverse topological order ensures that $\mathbf{f}_{e'}$ and $\mathbf{c}_{e'}$ are known for $e'\in I(v)$ before node $v$ is visited. 

\subsection{An Example}
We will now present an example of this algorithm applied to a sample solution for the modified butterfly network (Fig. \ref{butterfly}). Consider the following solution for the system over $GF(4)=\{0,1,\alpha,\alpha^2\}$, $\alpha^2=1+\alpha$.
\begin{align}
&a_1=a_5=b_2=b_6=1\nonumber\\
&a_2=a_6=b_1=b_5=0\nonumber\\
&a_3=a_4=\alpha\nonumber,\; b_3=b_4=\alpha ^2
\end{align}

One reverse topological order of edges is $1-2-3-4-5-6-7-8-9-10$. Nodes 1,2 are source nodes. So, we have $\mathbf{f}_{e_1}=\mathbf{f}_{e_4}=[1\ 0]$, $\mathbf{f}_{e_2}=\mathbf{f}_{e_5}=[0\ 1]$ and from the solution above, we have $\mathbf{c}_{e_1}=[0\ \alpha\ 1\ 0]$, $\mathbf{c}_{e_4}=[1\ \alpha]$, $\mathbf{c}_{e_2}=[0\ 1\ \alpha ^2\ 0]$, $\mathbf{c}_{e_5}=[\alpha ^2\ 1]$.

Then, beginning with the iteration for the non-source non-sink nodes as described in the algorithm above, we will first process node 3 - we know $\mathbf{f}_e,\mathbf{c}_e$ for both its input links, $e_1, e_2$. There are 4 copies of this node as shown in Fig. \ref{butterfly}b and all 4 copies have copies of edge $e_3$ as their only output links. Hence, $\mathbf{c}_{e_1,e_3}=\mathbf{c}_{e_1}$ and $\mathbf{c}_{e_2,e_3}=\mathbf{c}_{e_2}$. After computing $F_{e_1,e_3}$, $F_{e_2,e_3}$, we arrive at:
\begin{align*}
F_{e_3}= \left( \begin{array}{cccc}
0 & \alpha & 1 & 0\\
0 & 1 & \alpha ^2 & 0\end{array} \right)^T
\end{align*}

Now, let $\mathbf{f}_{e_3}=[\alpha\ 1]$, the second row of $F_{e_3}$. Hence, we have $a_{e_1,e_3}=\alpha, a_{e_2,e_3}=1$, the network coding coefficients at node 3. Also, $\mathbf{c}_{e_3}=[0\ 1\ \alpha ^2\ 0]$.

Next, we will move to node 4 which has two output links, $e_6,e_7$. Hence, we have $\mathbf{c}_{e_3,e_6}=[0\ 1], \mathbf{c}_{e_3,e_7}=[\alpha ^2\ 0]$. We can then compute $F_{e_3,e_6}, F_{e_3,e_7}$ and then arrive at:
$$
F_{e_6}= \left( \begin{array}{cc}
0 & 0\\
\alpha & 1\end{array} \right),\;\;F_{e_7}= \left( \begin{array}{cc}
1 & \alpha ^2\\
0 & 0\end{array} \right)
$$

Now we can choose $\mathbf{f}_{e_6}=[\alpha\ 1]$, the second row of $F_{e_6}$ and $\mathbf{f}_{e_7}=[1\ \alpha ^2]$, the first row of $F_{e_7}$. Then, the network coding coefficients for node 4 are $a_{e_3,e_6}=1,a_{e_3,e_7}=\alpha ^2$. Completing the last step of the iteration, we get $\mathbf{c}_{e_6}=[0\ 1], \mathbf{c}_{e_7}=[1\ 0]$.

Now we come to node 5. For the output link $e_8$, we have $\mathbf{c}_{e_4,e_8}=[1], \mathbf{c}_{e_6,e_8}=[0]$ and for $e_9$, we have $\mathbf{c}_{e_4,e_9}=[\alpha], \mathbf{c}_{e_6,e_9}=[1]$. Then we get:
$$
F_{e_8}=[1\ 0],\;\;F_{e_9}=[0\ 1]
$$

So, $\mathbf{f}_{e_8}=[1\ 0]$ and $\mathbf{f}_{e_9}=[0\ 1]$, the only rows of the respective matrices. Then, the network coding coefficients for node 5 are:
$$
a_{e_4,e_8}=1, a_{e_6,e_8}=0, a_{e_4,e_9}=\alpha, a_{e_6,e_9}=1
$$

Also, $\mathbf{c}_{e_8}=\mathbf{c}_{e_9}=[1]$.

Similarly, the network coding coefficients for node 6 can also be computed so that sinks 9, 10 receive the required symbols.

\section{Conclusion}
\label{sec:conclusion}
In this work, we have used path gains as variables to arrive at an algebraic formulation for the scalar linear network coding problem. This provides a useful simplification of the edge-to-edge gain formulation proposed in \cite{algebraic}, as illustrated by both small and large-sized examples. Given a network coding problem, we have given algorithms to construct an equivalent transformed network and arrive at a system of polynomial equations (of maximum degree 2) in terms of path gains. After solving for the path gains, we have provided an algorithm to compute the edge-to-edge gains, which can be used in implementing the network code.

Each monomial term occuring in a general system of polynomial equations can be assigned a new variable to obtain linear equations along with consistency conditions involving the new variables. However, in a general polynomial system, the consistency conditions are not guaranteed to be degree-2 equations without introducing additional monomial terms not present in the original system. Through this work, we have shown that the polynomial system representing a scalar network coding problem reduces to only degree-2 consistency conditions. 

\section*{Acknowledgements}
We thank the anonymous reviewers and the editor for their comments and suggestions that resulted in several significant improvements in the presentation and content of this article.
\newpage
\bibliographystyle{IEEE}

\vspace{-0.6in}
\begin{IEEEbiographynophoto}{Abhay T. Subramanian} obtained his bachelor's and master's degrees in Electrical Engineering from the Indian Institute of Technology (IIT) Madras, Chennai, India in 2008. He is currently a doctoral student at the Department of Management Science and Engineering in Stanford University, CA, USA. His current research interests are in the areas of applied probability, stochastic processes, optimization, statistics, algorithms, learning and mathematical finance.
\end{IEEEbiographynophoto}

\vspace{-0.6in}
\begin{IEEEbiographynophoto}{Andrew Thangaraj} received his bachelor's degree in Electrical Engineering from the Indian Institute of Technology (IIT) Madras, Chennai, India in 1998 and a PhD in Electrical Engineering from the Georgia Institute of Technology, Atlanta, USA in 2003. He was a post-doctoral researcher at the GTL-CNRS Telecom lab at Georgia Tech Lorraine, Metz, France from August 2003 to May 2004. From June 2004, he is with the Department of Electrical Engineering, IIT Madras, where he is currently an associate professor. His research interests are in the areas of information theory, coding and information-theoretic aspects of cryptography.
\end{IEEEbiographynophoto}
\end{document}